\documentstyle[psfig,draft]{mn}
\def\gtsima{$\, \buildrel > \over \sim \,$}
\def\ltsima{$\, \buildrel < \over \sim \,$}
\def\simgt{\lower.5ex\hbox{\gtsima}}
\def\simlt{\lower.5ex\hbox{\ltsima}}
\def\sm{$\sim\,$}

\def\nhat{\ifmmode {\hat{\bf n}}\else${\hat {\bf n}}$\fi}

\def\degs{\ifmmode^\circ\else$^\circ$\fi}
\def\kms{\ifmmode{\rm km}\,{\rm s}^{-1}\else km$\,$s$^{-1}$\fi}

\def\etal{{\sl et al.}}
\def\apriori{{\rm a priori}}

\def\h1{\ifmmode h^{-1}\!\else$h^{-1}\!$\fi}
\def\msun{M_\odot}

\def\omegam{\Omega_{\rm M}}
\def\omegal{\Omega_\Lambda}
\newcount\itemno
\def \ino         { \the\itemno\global\advance\itemno by 1 }

\title[]
{Weak Lensing Constraints on Cluster Structural Parameters}
\author[Willick \& Padmanabhan] 
{Jeffrey A.\ Willick$^{1}$ \& Nikhil Padmanabhan \\ 
Department of Physics, Stanford University, Stanford, CA 94305-4060 \\
 E-mail: npaddy@stanford.edu \\
 $^{1}$ Deceased}
\date{\today}
\begin{document}
\maketitle
\begin{abstract}

%auto-ignore

We study the reliability of dark matter (DM)
 profile determinations
for rich clusters of galaxies based on weak gravitational
lensing measurements. We assume that cluster
DM density profiles are fully described by two parameters:
a scale radius $r_s$ and a dimensionless central density $\delta_c.$
We consider three functional forms for the mass distribution:
a softened isothermal sphere (SIS), and two models proposed
on the basis of numerical simulations of CDM universes: the
Navarro, Frenk, \& White  model (NFW) and the Klypin, Kravstov,
Bullock, \& Primack model (KKBP). A maximum likelihood parametric
method is developed and applied to 
simulated images of background galaxies at z=2.0 lensed
by clusters at $z=0.2,$ $0.5,$ and $0.8.$
The images have field size ($\sim 10$ arcmin) and angular
resolution ($\sim 0.5$ arcsec) characteristic of the new generation
of 8--10 meter-class ground-based telescopes. 
The method recovers the correct virial mass of the 
DM halo, $M_{200},$
to within $\sim 10\%$ at all redshifts. 
The recovered values of $\delta_c$ and
$r_s$ are only accurate to \sm 25\%; however, their correlated 
errors are such that greater accuracy is obtained for the mass. 
Clusters with an SIS density profile are recognized
by the method, while NFW and KKBP clusters
are recognized as not being SIS. The method does
not discriminate between NFW and KKBP profiles
for clusters of a given mass, because they are too similar
in their outer regions. Strongly lensed arcs
near the cluster centers would be needed to distinguish
NFW from KKBP. The accuracy with which
$\delta_c$ and $r_s$ are determined, and the
different models are distinguished, {\em improves\/}
with increasing redshift because the
gravitational potential is sampled out to a larger
radius for a given field size. These preliminary results 
conclude that our maximum likelihood method is a promising  
candidate for both virial mass measurements and testing numerical 
predictions of DM halo structures although further tests that include 
real world effects still remain to be done.

\end{abstract}

\section{Introduction}
\label{sec:intro}
%auto-ignore

The study of rich clusters of
galaxies, over a range of redshifts out to $z\sim 1,$ has emerged
as an important tool for studying structure formation
and constraining cosmological parameters 
(White, Efstathiou, \& Frenk 1993; Bahcall \& Cen 1993; 
Carlberg \etal\ 1997; Bahcall, Fan, \& Cen 1997; 
Riechart \etal\ 1998; Eke \etal\ 1998; Girardi \etal\ 1998; 
Borgani \etal\ 1999; Blanchard \etal\ 2000).
The theoretical underpinning of this approach
is the Press \& Schechter (1974; PS) 
approximation for the abundance of collapsed objects as
a function of mass and redshift in a given
cosmological model. The PS formalism
is extremely sensitive to cluster virial mass
since the fraction
of collapsed objects of mass $> M$ is given by the integral of the tail
of a Gaussian function with the lower limit of integration
dependent on $M$ and cosmological parameters. 
Therefore, even relatively modest
errors in virial mass measurement can drastically affect the derived
cosmological parameters (Willick 2000). Accurate virial mass
determination is thus of paramount importance.
 
Observational data for clusters do not, however, directly yield their masses. 
Instead, they measure
related properties such as X-ray luminosity and/or temperature, 
cluster richness, or velocity
dispersion. 
To date, efforts to constrain cosmological parameters using clusters
have been based mainly on either X-ray temperature/luminosity 
(Girardi \etal\ 1998; Donahue \etal\ 1998; Sadat, Blanchard,
\& Oukbir 1998; Riechart \etal\ 1998; Borgani \etal\ 1999;
Blanchard \etal\ 1999),
or  velocity dispersion
(Carlberg \etal\ 1998; Tran \etal\ 1999) measurements as mass indicators. 
However, the accuracy of such mass estimates
is open to question.
They depend on assumptions of
hydrostatic equilibrium of the hot intracluster
gas and virial equilibrium of cluster galaxies, respectively, 
which may not hold
in a given cluster. It is thus
important that alternative mass estimators, independent of the assumption of
hydrostatic or virial equilibrium, also be applied when possible.

Weak gravitational lensing of background sources by clusters is
perhaps the most attractive such alternative
(see Narayan \& Bartelmann 1999, hereafter NB99, and Bartelmann \&
Schneider 1999, hereafter BS99, for recent reviews). While strong
lensing, in which large arcs or multiple images of a single
source are produced, can provide precise measurements
of the mass in the inner regions ($r \simlt 200\h1$ kpc)
of clusters (e.g., Tyson, 
Kochanski, \& Dell'Antonio 1998), it does not constrain
the virial mass of clusters, which typically corresponds
to radii of 1--2\h1\ Mpc. 
In contrast, weak lensing, which consists
of small but coherent distortions of background galaxy shapes,
can be detected statistically out to several \h1\ Mpc.
Cluster mass estimates 
have now been obtained from weak lensing data  by
a number of groups (e.g., Tyson, Valdes, \& Wenk 1990;
Kaiser \& Squires 1993, 1996; Smail \etal\ 1997; Squires \etal\ 1997;
Hoekstra \etal\ 1998;
Luppino \& Kaiser 1997). These estimates have generally
been found to agree well
with the X-ray
and velocity dispersion mass estimates (Lewis \etal\ 1999;
Willick 2000).

Several of the these weak lensing
studies were done with the Hubble Space Telescope
(HST) whose superb imaging capabilities
ensure that it will play an important
role in future projects as well. However,
the primary observational driver for future
work is likely to be improvements
in the capabilities of the new
generation of large aperture ground-based 
telescopes. Wide field
($\sim 10'$--1\degs) imaging CCD cameras 
have recently or will soon come on line
(see, e.g., Luppino, Tonry, \& Stubbs 1998; 
Kaiser, Tonry, \& Luppino 2000) at
telescopes with excellent image quality ($\simlt 0.5''$ seeing),
such as CFHT, Subaru, Magellan, VLT, and Gemini. 
These instruments are extremely well-suited
to weak-lensing studies. In particular, they will fully
cover the field of moderate-redshift ($0.2 \simlt z \simlt 1$)
clusters, with data that probe
out to and beyond the virial radius.

The problem of cluster mass ``inversion'', i.e. reconstruction
of the mass as a function of position from weak lensing data, has
received considerable attention 
in recent years (e.g., Kaiser \& Squires 1993; 
Seitz \& Schneider 1997; Bartelmann 1998).
Most of these efforts have aimed at 
nonparametric mass reconstructions, in which no \apriori\ 
assumptions are made about the underlying mass distribution. However,
there is 
evidence from numerical simulations 
that the dark matter distribution
in virialized haloes might be well described by simple parametric
models. Navarro, Frenk, \& White (1997; NFW) have proposed
a model (\S 2) that has received perhaps the most attention. Recent
high-resolution simulations have suggested departures
from the NFW model at small radii, leading Kravstov, Klypin, Bullock, \&
Primack (1998; KKBP) to propose an alternative parameterization.
Recent efforts to distinguish between these alternatives using the
rotation curves of dwarf galaxies have been inconclusive
(van den Bosch \& Swaters, 2000), leaving it
an open problem.
The NFW and KKBP distributions
are, however, similar at radii encompassing $\simgt 10\%$ of the virial
mass, where both may be described as 
locally power-law profiles $\rho(r) \propto r^{-\gamma},$
where $\gamma$ varies slowly from $\simlt 2$ near
a characteristic radius $r_s$ to
$3$ at $r\gg r_s.$ In this sense, NFW and KKBP differ distinctly
from the generic softened isothermal sphere (SIS) model, 
in which $\rho(r) \propto r^{-2}$
at all radii beyond a small inner core. The SIS model yields
the flat rotation curves characteristic of spiral galaxies,
and thus has often been viewed as a reasonable representation
of equilibrium dark halos. 

Given the findings by NFW and KKBP that
dissipationless gravitational clustering may
produce simple and universal
mass distribution laws, a strong case can be made for
using weak lensing to constrain the parameters of
such laws. This approach would complement the
existing methods of nonparametric mass inversion.
Fitting parametric models has advantages
and disadvantages. The chief advantage is that, if
the model is correct or nearly so, a parametric approach
is robust against random error in
the presence of noise. The obvious disadvantage is
that the method may not flag an incorrect
model. Nonetheless, we believe that the
case for universal density profiles is strong enough
that it amounts to a ``theoretical prediction'' which
can and should be fit to the data.

This paper has two objectives,
to see if we can distinguish between several
theoretically motivated density profiles using
weak lensing data; and second, to ascertain
how accurately we can measure virial masses from such data.
We adopt three
trial parameterizations of cluster masses: SIS, NFW, and
KKBP (see \S~2 for details), and simulate lensed fields
for each of these mass distributions, for clusters of
varying mass and redshift. We then apply our 
maximum likelihood method to these
simulations (\S~3). For each cluster of a given profile type, we
compute the likelihood with respect to the parameters
of that profile type {\em and of the other two,} allowing
us not only to constrain the parameters but to test
whether we can discriminate among the models. One of our
key conclusions is that we can indeed distinguish
the SIS from NFW or KKBP, but not
NFW from KKBP.
This is important, because a basic prediction of the
Cold Dark Matter (CDM) hypothesis is that halo mass distributions
depart from an $r^{-2}$ law in their outer regions. Our
work shows that this prediction is testable via
weak lensing data.

In addition to developing a method 
to test theoretical predictions of
cluster mass profiles, our work has another, 
more basic, goal, already alluded to
in the first paragraph. For clusters to be
used as cosmological probes via the PS or
related formalisms, one must determine their
{\em virial masses\/} with high accuracy.
In the PS formalism, the term virial mass has
a very specific meaning: it is the mass contained
within a sphere whose mean interior density
is a particular multiple, $\delta_V,$ of the
mean density of the universe. The quantity
$\delta_V$ depends on cosmological parameters
and on redshift. Since the former are not
known \apriori---and indeed may be the
goal of our study---one needs to be able
to express mass as a function of radius
in a flexible way. 
This cannot be done independently of a
model for the density profile of the cluster;
moreover, once a model is adopted, its defining parameters
must be reasonably well-known to obtain a virial
mass for a given choice of cosmological parameters
(see, e.g., Willick 2000 for a detailed discussion).

The outline of this paper is as follows: Section 2 and 3 introduce
the basic lensing formalism and our parametric
maximum likelihood inversion method. We then discuss the simulations 
used to test the method in Section 4, and Section 5 presents the results 
of these simulations. Section 6 discusses the basic conclusions and 
directions for further research. A derivation of the shear and magnification 
for SIS, NFW and KKBP clusters is in Appendix 1, whereas Appendix 2 considers
the computation of the virial mass for a cluster with known structural 
parameters.

\section{Basic Formalism}
\label{sec:formalism}
%auto-ignore

\def\ys{y_{_S}}
\def\yl{y_{_L}}
\def\yls{y_{_{LS}}}
\def\zs{z_{_S}}
\def\zl{z_{_L}}
In this section we derive expressions for lensing
angle, convergence, and shear for two-parameter,
spherically symmetric mass
distributions such as SIS, NFW, and KKBP. 
Such mass profiles are characterized by
a density scale $\rho_c$ and a scale length $r_s.$\footnote{The
quotation marks are meant to indicate that $\rho_c$
is not necessarily the density at $r=0,$ which may in fact be infinite,
but rather is a characteristic density in the inner
regions of the profile. Its precise meaning depends
on the mathematical form of the profile.}
We follow NFW in writing $\rho_c=\rho_{crit}\,\delta_c,$ where
$\rho_{crit}=3H_0^2/8\pi G,$ and thus 
write the density law as
\begin{equation}
\rho(r) = \rho_{crit}\,\delta_c\,\eta(r/r_s)\,,
\label{eq:dens_profile}
\end{equation}
where the function $\eta(x)$ characterizes the shape of the density profile.
For the three mass distributions
considered in this paper we have
\begin{equation}
\eta(x) = \left\{ \begin{array}{ll}
		(1+x^2)^{-1} & {\rm (SIS)} \\
		x^{-1}(1+x)^{-2} & {\rm (NFW)} \\
		x^{-0.2} (1+x^2)^{-1.4} & {\rm (KKBP).}
	        \end{array} \right.
\label{eq:mprof}
\end{equation} 

\begin{figure}
\psfig{figure=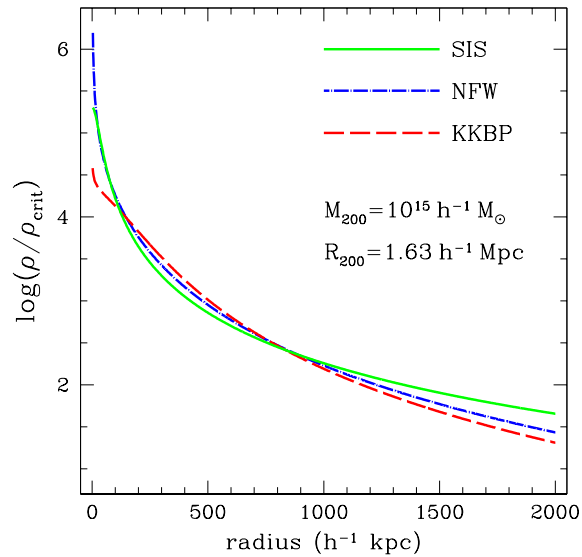,height=8truecm,width=8truecm}
\caption{%
Mass density, in units of the critical
density, plotted as a function of radius
for three clusters each with $M_{200}=10^{15}\h1\msun.$
One cluster has an SIS density profile, one
has an NFW profile, and one has a KKBP profile.
The cluster structural parameters $r_s$ and
$\delta_c$ are chosen to make the three clusters
as similar as possible. See text for further details.
}
\label{fig:compdens}
\end{figure}

Figure \ref{fig:compdens} shows these three profiles
for the case of a cluster with virial mass $M_{200} = 10^{15} 
h^{-1} M_{\odot}$
and virial radius 
$R_{200} = 1.63 h^{-1} Mpc$, for 
$\delta_{v} = 200$. Spherical symmetry
implies circular symmetry on the sky. The
lensing angle, $\alpha(\theta)$ must be purely radial,
dependent only on $\theta$, the angle on
the sky from the center of the mass distribution. Furthermore circular
symmetry implies (e.g., NB99) that
\begin{equation}
\alpha(\theta) = \frac{4GM(\theta)}{c^2 \theta}\frac{D_{LS}}{D_L D_S}\,,
\label{eq:lensangle}
\end{equation}
where $M(\theta)$ is the mass enclosed within a circle of angular
radius $\theta,$
and the various $D$'s are angular diameter distances. The subscript
$L$ refers to the lens, i.e., the cluster, and the subscript $S$
to the source, i.e, the object being lensed.
For reference, we write down expressions for these 
distances as a function of redshift. We follow
the notation of Peebles (1993) for cosmological
quantities, except that we use the
subscript $M$ rather than $0$ to denote the mass density
parameter.
For $D_L$ ($D_S$)
we have 
\begin{equation}
D(z) = \frac{c}{H_0}\,\frac{y(z)}{1+z}
\end{equation}
where $z$ is the lens (source) redshift, and $y(z)$ is given,
in the case of a flat ($\omegam+\omegal=1$) universe, by
\begin{equation}
y(z)=
\int_0^z \frac{dz'}{\sqrt{\omegam(1+z')^3 +\omegal}}\,.
\label{eq:defyz}
\end{equation}
The lens-source angular size distance is  
\begin{equation}
D_{LS} = 
\frac{c}{H_0}\,\frac
{\yls}{1+\zs}\,,
\end{equation}
where we have introduced
the abbreviations $\yl=y(\zl),$ $\ys=y(\zs),$ 
and $\yls=\yl-\ys$ (again,
assuming a flat universe). Using
the above expressions, we have
\begin{equation}
\frac{D_{LS}}{D_L D_S} = \frac{H_0 (1+\zl)}{c}\,\frac{\yls}{\yl \ys}
\label{eq:dratio}
\end{equation}
This last form of the distance ratio will be useful in what follows.

Adopting the two-parameter density profile (Eq.~\ref{eq:dens_profile}),
the projected mass within
an angle $\theta$ on the sky is given by
\[ M(\theta) = 2\pi \int_0^{D_L\theta}\!r 
\,dr \int_{-\infty}^{\infty} \rho\left(\sqrt{r^2+z^2}\right) dz \]
\begin{equation}
= 4\pi\rho_{crit}\,\delta_c D_L^3 \theta_s^3 \int_0^{\theta/\theta_s} 
\! y\,dy \int_0^\infty
\eta\left(\sqrt{y^2+x^2}\right) dx\,,
\label{eq:mtheta}
\end{equation}
where 
\begin{equation}
\theta_s\equiv r_s/D_L
\end{equation}
is the halo scale length in angular units.
Combining Eqs. \ref{eq:lensangle}, \ref{eq:dratio}, and
\ref{eq:mtheta}, and using the definition of the critical
density, we obtain, after some algebra,
the deflection angle:
\begin{equation}
\alpha(\theta) = 6\,\delta_c\,\theta_s^2\, {\cal Y}(\zl,\zs) 
\left(\theta/\theta_s\right)^{-1} \beta
(\theta/\theta_s)\,, 
\label{eq:final_lens}
\end{equation}
where 
\begin{equation}
\beta(x) \equiv \int_0^x y\,dy \int_0^\infty 
\! \eta\left(\sqrt{y^2+w^2}\right) dw
\label{eq:defbeta}
\end{equation}
and
\begin{equation}
{\cal Y}(\zl,\zs) \equiv \left(\frac{\yls}{\ys}\right)
\left(\frac{\yl}{1+\zl}\right)^2\,.
\label{eq:defcaly}
\end{equation}
Eqs. \ref{eq:final_lens} and \ref{eq:defbeta} determine
the lensing angle for any adopted (spherically symmetric)
density law. Note that the lens and source redshifts enter
only through the 
multiplicative factor ${\cal Y}(\zl,\zs),$ while the
shape information is contained entirely in the function $\beta(x).$

\subsection{Convergence and Shear}
Our formula for $\alpha(\theta)$ enables us to calculate
two other important quantities, the convergence $\kappa$ and
shear $\gamma.$ To obtain the former, note that
\begin{equation}
\kappa = \frac{1}{2} \nabla\cdot\vec{\alpha}\,
=\, \frac{1}{2}\left[\frac{d\alpha}{d\theta}+\frac{\alpha}{\theta}\right]\,,
\label{eq:kappafromalpha}
\end{equation}
where the last equality follows from circular symmetry.
Using Eq.~\ref{eq:final_lens} in Eq.~\ref{eq:kappafromalpha}, 
we obtain
\begin{equation}
\kappa(x) = 3\delta_c\theta_s {\cal Y}(\zl,\zs) \left[x^{-1}\beta^\prime(x)
\right]\,,
\label{eq:kappa}
\end{equation}
where $x=\theta/\theta_s.$

The complex shear, $\gamma=\gamma_1+{\rm i}\gamma_2,$ is given by
\begin{equation}
\gamma_1 = \frac{1}{2}\left(\psi_{,xx}-\psi_{,yy}\right)
\end{equation}
and
\begin{equation}
\gamma_2 = \psi_{,xy}
\end{equation}
where $\psi(\theta)$ is the {\em lensing potential\/} (NB99)
defined by $\vec{\alpha}=\nabla\psi,$ and the comma
signifies partial differentiation. 
Circular symmetry implies
that $\alpha=\psi^\prime,$ giving, after some manipulation,
\begin{equation}
\gamma_1 = \frac{1}{2}\cos 2\phi\left[\frac{d\alpha}{d\theta} 
- \frac{\alpha}{\theta}\right]
\label{eq:gam1}
\end{equation}
and
\begin{equation}
\gamma_2 = \frac{1}{2}\sin 2\phi\left[\frac{d\alpha}{d\theta} - 
\frac{\alpha}{\theta}\right]
\label{eq:gam2}
\end{equation}
where $\phi$ is the angle between the radial direction
and the $x$-axis. The shear amplitude is
\[ |\gamma|=\sqrt{\gamma_1^2+\gamma_2^2}=(\alpha'-\alpha/\theta)/2 \]
\begin{equation}
= 3\delta_c\theta_s{\cal Y}(\zl,\zs) \left[x^{-1}\beta^\prime(x)
-2x^{-2}\beta(x)\right]\,,
\label{eq:shear_ampl}
\end{equation}
where as before $x=\theta/\theta_s$ and 
Eq.~\ref{eq:final_lens} has once again been used. 

Note that Eqs.~\ref{eq:gam1} and~\ref{eq:gam2} imply that
for $\phi=0$ (i.e., a point along the $x$-axis),
$\gamma_1=|\gamma|$ and $\gamma_2=0.$ However,
because of the circular symmetry of the problem,
there is no natural orientation for the two
Cartesian axes. Thus, we may reinterpret the above equations
as saying that at any point, we may
adopt a local Cartesian coordinate system in which
one axis points along the radial direction, the
other along the tangential direction. With respect
to such a system, the real part of $\gamma$ is
always equal to $|\gamma|$ as given by Eq.~\ref{eq:shear_ampl},
while the imaginary part vanishes. 
As described below, we can measure complex galaxy ellipticities
relative to such local axes. The resulting simplification
in the probability distribution of observed ellipticities
is described in \S~\ref{sec:method}.

\subsection{Discussion}

The formalism described above shows that the relevant
quantities for a lensing analysis---the bending angle $\alpha,$
the convergence $\kappa,$ and the shear $\gamma$---are
determined (up to cosmological factors) by three quantities: the
dimensionless central density $\delta_c,$ the scale radius $r_s,$
and the dimensionless projected mass function $\beta(x)$ given
by Eq.~\ref{eq:defbeta}. The last quantity depends only
on the assumed form of the mass distribution (SIS, NFW, KKBP,
or other); the former two may then be viewed as parameters
to be determined by the lensing analysis. In Appendix~1,
we derive expressions for $\beta(x)$ for SIS, NFW, and KKBP.

\begin{figure}
\psfig{figure=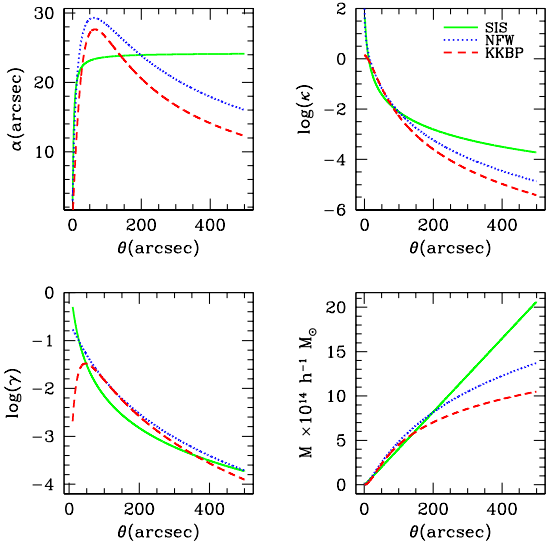,height=8truecm,width=8truecm}
\caption{%
Lensing angle $\alpha,$ convergence $\kappa,$ shear
amplitude $\gamma,$ and enclosed mass $M$
plotted as functions of $\theta,$ the angle on 
the sky from the center of the mass distribution, for
a cluster of virial mass $10^{15} \h1\msun$ at
a redshift of $0.5.$ Each quantity is shown for
each of the three density profiles studied in
this paper: SIS, NFW, and KKBP. The three profiles
were ``matched'' as described in the main text. 
Note that the parameters are the same as those in Figure
\ref{fig:compdens}.
}
\label{fig:models}
\end{figure}

In Figure~\ref{fig:models}, we compare the resultant forms
of $\alpha(\theta),$ $\kappa(\theta),$ $\gamma(\theta),$
and $M(\theta)$ for a fiducial cluster of virial mass $10^{15}\h1\msun$
lying at $z=0.5.$ The parameters $\delta_c \approx 27000 $ and $r_s=200\h1$ kpc
were adopted for the NFW profile, while for
SIS we took the core radius to be $10\h1$ kpc. The parameters for KKBP 
were determined by requiring that the logarithmic slope
of the density profile, $d\ln\rho/d\ln r,$ take on
the value $-2$ at $r=200\h1$ kpc, where the NFW logarithmic
slope has that value as well. These conditions plus 
fixing the virial mass at $10^{15}\h1\msun$ uniquely determines 
the profiles and makes the comparison well-defined.

\section{Maximum-Likelihood Method for Determining Halo Parameters}
\label{sec:method}
%auto-ignore

\subsection{Preliminaries}
\def\beps{\mbox{\boldmath$\epsilon$}}
The method is based on
the (complex) ellipticities, $\beps,$ of galaxy images, 
computed from
the intensity moments $Q_{ab}$,
\begin{equation}
\beps = \epsilon_{x}+i\epsilon_{y} = \frac{Q_{xx}- Q_{yy} + 2iQ_{xy}}{Q_{xx}+Q_{yy}}
\end{equation}
where
\begin{equation}
Q_{ab} = \frac{\int I\,a\,b\,dy\,dx}{\int I\,dy\,dx} \,,
\end{equation}
and $I$ is the measured intensity.
As noted above, the
formalism is considerably simplified if we transform 
to a coordinate system in which the real axis
(axis 1) points along the tangential direction, 
and the imaginary axis (axis 2) along the radial direction, with respect
to a radial vector joining the lens center and the galaxy.
This transformation is accomplished by a simple rotation,
\begin{equation}
\left(\!\begin{array}{c}
\epsilon_1\\
\epsilon_2
\end{array}\!\right)
= \left (\! \begin{array}{rr}
		\sin 2\phi  & \cos 2\phi  \\
		-\cos 2\phi  & \sin 2\phi  
	    \end{array}\! \right )
\left(\!\begin{array}{c}
		\epsilon_x \\
		\epsilon_y 
\end{array}\!\right)
\label{eq:trsfm}
\end{equation}
where $\phi$ is the angle between the original $x$-axis and
the real axis at the position of the galaxy.

The ellipticity components $\epsilon_1$ and $\epsilon_2$ transform under
lensing in a particularly simple way. As shown by Miralda-Escud\'{e} (1991,
hereafter ME91), the ellipticity components of the lensed image
(subscript ``$L$'') are related to those of the source
image (subscript ``$S$'') by the transformations
\begin{equation}
\epsilon_{L1}= \frac{\delta +\epsilon_{S1}}{1+\delta\epsilon_{S1}}
\label{eq:ls1}
\end{equation}
and
\begin{equation}
\epsilon_{L2}=\frac{(1-\delta^{2})^{1/2}\epsilon_{S2}}{1+\delta\epsilon_{S1}}\,.
\label{eq:ls2}
\end{equation}
In Eqs.~\ref{eq:ls1} and~\ref{eq:ls2}, $\delta$ is the
{\em distortion\/} defined by (e.g., ME91)
\begin{equation}
\delta=\frac{1-q_{c}^{2}}{1+q_{c}^{2}}\,,
\end{equation}
where
\begin{equation}
q_{c}=\frac{1-(\kappa+\gamma)}{1-(\kappa-\gamma)}
\end{equation}
and $\kappa$ and $\gamma$ are the convergence and shear amplitude,
respectively. % which are calculated as described above.

Our method requires that we invert these transformations
to obtain the source ellipticity as a function of the
image (lensed) ellipticity. This yields
\begin{equation}
\epsilon_{S1}= \frac{-\delta +\epsilon_{L1}}{1-\delta\epsilon_{L1}}
\label{eq:invsl1}
\end{equation}
\begin{equation}
\epsilon_{S2}=\frac{(1-\delta^{2})^{1/2}\epsilon_{L2}}{1-\delta\epsilon_{L1}}
\,.
\label{eq:invsl2}
\end{equation}
The Jacobian of the inverse transformation is given by
\begin{equation}
{\cal J}=\left (\frac{\sqrt{1-\delta^{2}}}{1-\delta\epsilon_{L1}}\right )^{3}
\,.
\label{eq:jacob}
\end{equation}

\def\epone{\epsilon_1}
\def\ep2{\epsilon_2}
\def\eps{\epsilon_S}
\subsection{Probability Distributions}
We seek an expression for the probability that the image of a background galaxy
will exhibit ellipticity components $(\epone,\ep2)$ (we drop
the subscript $L$ in what follows). Following the work of Seitz \&
Schneider (1995,1996; see also Schneider \& Seitz 1995),
we write the probability density in terms of the image
ellipticity components as
\begin{equation} 
p(\epone,\ep2;\delta) = \phi\!\left[\eps(\epone,\ep2;\delta)\right] 
\times {\cal J}(\epone;\delta)\,, 
\label{eq:probeps}
\end{equation}
where $\eps\equiv \sqrt{\epsilon_{S1}^2+\epsilon_{S2}^2},$
and $\epsilon_{S1}$ and $\epsilon_{S2}$
are calculated from the inverse
transformation, Eqs.~\ref{eq:invsl1} and~\ref{eq:invsl2}.
The function $\phi(\eps)$ is the distribution of intrinsic
source ellipticities; it only depends on
amplitude because of the assumed isotropy of
source galaxy orientations, and is normalized by
\begin{equation}
2\pi\int_0^1 d\epsilon\, \phi(\epsilon) = 1\,.
\end{equation}
 A successful application of the maximum-likelihood method
requires an accurate model for $\phi(\epsilon).$ We discuss
this further in \S~6.

The data set with which we maximize likelihood consists of
$N$ lensed galaxy images, each with
a measured complex ellipticity $\beps_i.$ 
We define the likelihood for the data set as
\begin{equation}
{\cal L} = - \sum_{i=1}^{N} 
\log p(\beps_i;\delta_i(\delta_{c}, \theta_{s}),\theta_i)
\label{eq:like}
\end{equation}
where the individual probabilities are given by
Eq.~\ref{eq:probeps}. Here we have explicitly
indicated the dependence of the probability on
$\theta_i,$ the angle between the radius vector   
connecting the cluster center to the galaxy and
the $x$-axis,
which is needed for calculating 
$\epone$ and $\ep2$ from the
directly measured components $\epsilon_x$ and $\epsilon_y$
(cf.\ Eq.~\ref{eq:trsfm}). The likelihood is maximized
with respect to the cluster structural parameters 
$\delta_c$ and $\theta_s=r_s/D_L$.

\section{Simulating Lensed Background Fields}
\label{sec:simulate}
%auto-ignore

%Outline:
%
%1. Intro para
%2. Description of simulations
%	2a. Types of fields lensed -x
%	2b. Lensing clusters - their parameters -x
%		2b(i) Determination of parameters -x
%	2c. Method of lensing used
%	2d. Image processing
%	2e. Parameter processing
%
%

In order to quantify the feasibility of the described method, we applied it to  
synthetic data sets, that covered a wide area in parameter space. 
These simulations are described in this section; the results are presented in the 
following section.

\begin{figure}
\psfig{figure=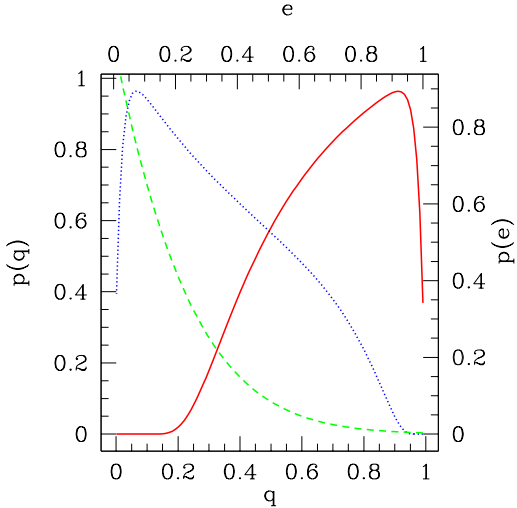,height=7truecm,width=7truecm}
\caption{%
The intrinsic ellipticity distribution. The solid line shows $p(q)$
(Eq. \ref{eq:e_dist})
, the dotted line shows 
the corresponding
$p(e)$ while the dashed line shows the observed distribution
$p_{obs}(e)$ (Eq. \ref{eq:e_dist_real}). An arbitrary normalization
is chosen. The effects of pixelization, increasing the number of 
galaxies with low ellipticities is also evident.}
\label{fig:ellip}
\end{figure}

\begin{figure}
\psfig{figure=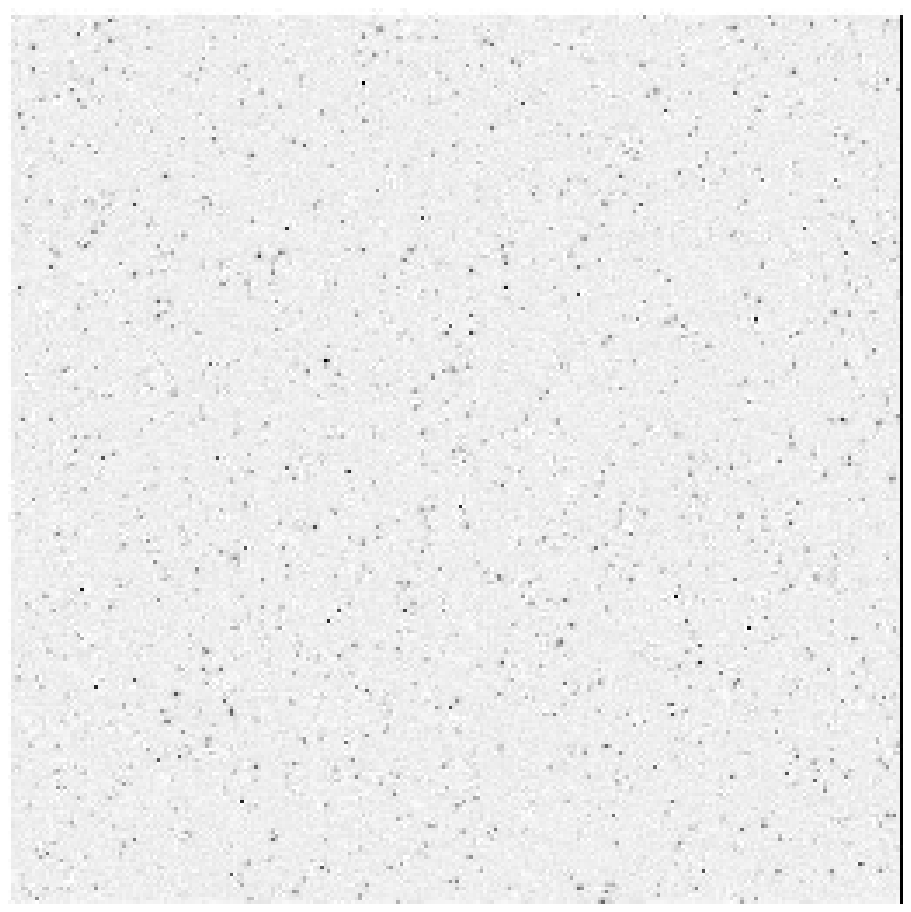,height=7truecm,width=7truecm}
\caption{%
A sample field of background galaxies, seen as they
would be in the complete absence of gravitational lensing.
}
\label{fig:field}
\end{figure}

The simulations involved generating $10$ fields of galaxies, $10'$ on a side with 
a resolution of $0.3''$ per pixel. Each field  had approximately $1500$ galaxies 
randomly distributed per field with an intrinsic ellipticity distribution given 
by ME91
\[ p(q)dq \propto (1+cq) \exp\left[-\left(\frac{q_1}{q}\right)^2\right] \]
\begin{equation}
\times \left[1-\exp\left(-\frac{1-q}{q_2}\right)\right] dq \,,
\label{eq:e_dist}
\end{equation}
where $q$ is the ratio of minor to major axes,\ $q=b/a$ and with 
$c=0.2$,$\, q_1=0.4$ and $q_2=0.02$. This probability distribution 
is shown in Figure \ref{fig:ellip}. The choice 
of this functional form and the 
particular parameters for the ellipticity distribution 
is a best fit
(ME91)
to the ellipticity distribution of the deep CCD survey (Tyson 1988, 
Tyson \& Seitzer 1988) while the number of galaxies was chosen such that the 
number of galaxies per unit area is characteristic of real images.
Each of these $10$ fields were assumed to be 
at a redshift\ $z=2.0$ in a \ 
$\Omega_m=0.3$,$\, \Omega_\Lambda=0.7,h = 0.7$  universe. 
We added uniform Poisson photon noise to each of the images, but ignored 
the effects of a redshift distribution and seeing for simplicity. We realize that
the latter should be included at a later stage.
An example of a
background field is shown in Figure \ref{fig:field}. 

Each of these fields was lensed by dark matter haloes of virial
masses $5.0$,$\, 7.5$, 
$10.0$ and $15.0 \times 10^{14} M_\odot$ each located at redshifts of 
$z=0.2$,$\, 0.5$ and $0.8$. For each halo of a given mass, we simulated three 
different density profiles, SIS, NFW and KKBP, 
to study variations 
with mass, redshift, and profiles. The choice of the particular  
profiles enabled us to test the algorithm's discriminatory power at varying 
scales since the KKBP profile differs from NFW only at small radii while both 
differ from SIS at all length scales.

However, since all the profiles are two parameter models, specifying the mass 
does not uniquely determine the values of both parameters. Recognizing that the 
halo scale length (for NFW and KKBP) and the cluster core radius (SIS) are 
parameters that lack significant physical significance, 
we fixed those 
parameters in our simulations. The scale lengths for SIS and NFW were 
arbitrarily set to $10$ kpc and $200$ kpc respectively. To determine 
the scale length for KKBP, we matched the radius at which $d\log \rho / d\log r = 
-2$ with that for NFW; this fixed the $r_{KKBP}=r_{NFW}/\sqrt{1.8}=149.07$ kpc. 
This determined the second parameter, the overdensity $\delta_c$ or 
the velocity dispersion $\sigma_v$, for a halo of a given mass uniquely. 

\begin{figure}
\psfig{figure=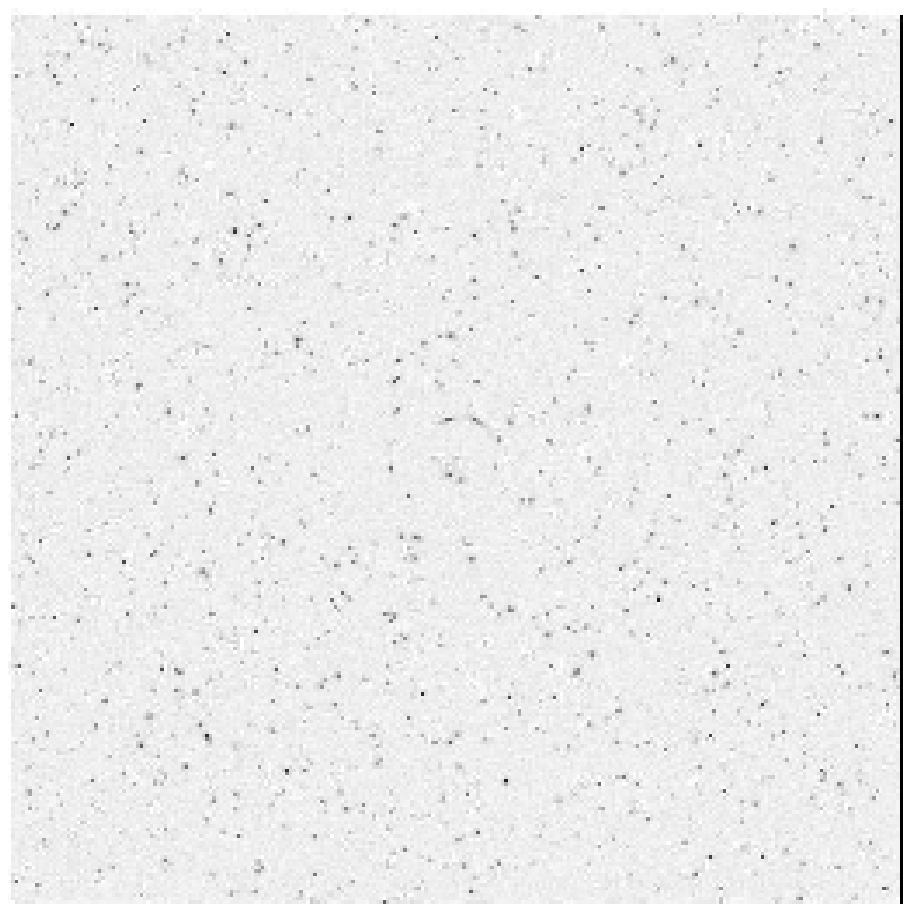,height=7truecm,width=7truecm}

\caption{%
The same field lensed with an SIS halo of mass $m=1.0 \times 10^{15} M_\odot$ at 
$z=0.5.$
}
\label{fig:fieldsis}
\end{figure}

\begin{figure}
\psfig{figure=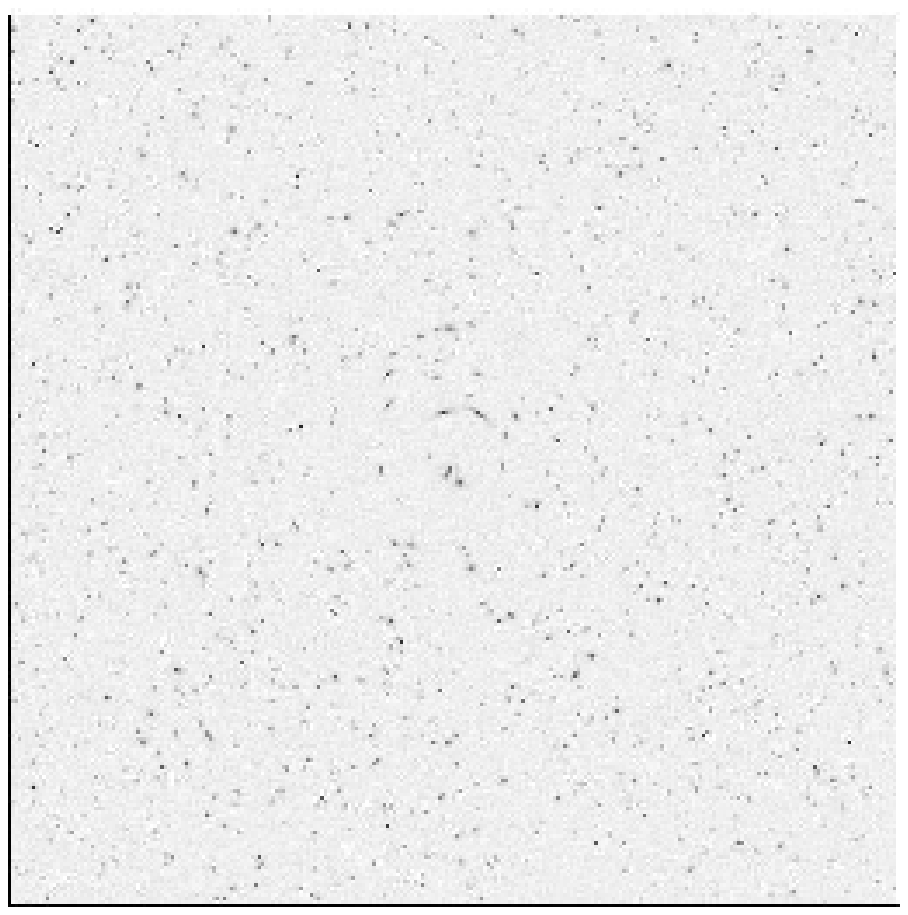,height=7truecm,width=7truecm}
\caption{%
The same field lensed with an NFW halo of mass $m=1.0 \times 10^{15} M_\odot$ at 
$z=0.5.$
}
\label{fig:fieldnfw}
\end{figure}

\begin{figure}
\psfig{figure=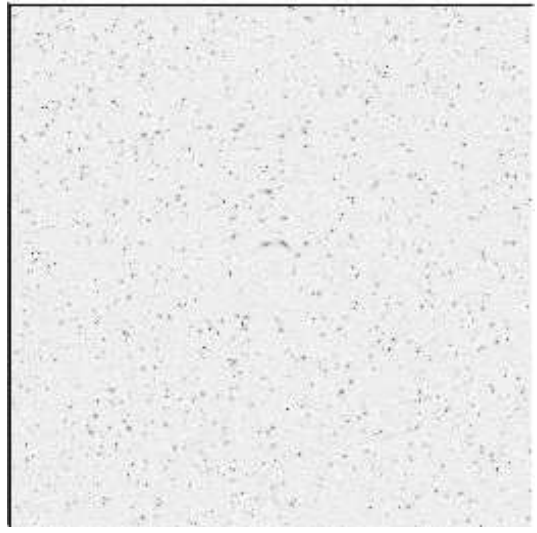,height=7truecm,width=7truecm}
\caption{%
The same field lensed with an KKBP halo of mass $m=1.0 \times 10^{15} M_\odot$ 
at $z=0.5.$
}
\label{fig:fieldkkbp}
\end{figure}

The actual lensing used an algorithm described 
in Wilson et al.(1996); it involved solving the lens equation
\begin{equation}
\beta=\theta - \alpha(\theta)
\end{equation}
where $\beta$ is the radial position of the unlensed image, 
$\theta$ is the radial position of the lensed image and 
$\alpha(\theta)$ is the lensing angle, for every point on the 
\textit{image} frame and mapping it to the \textit{source} frame, thereby 
correctly handling multiply imaged sources.
Examples of these lensed images are shown in Figures \ref{fig:fieldsis},
\ref{fig:fieldnfw} and \ref{fig:fieldkkbp}.
The frames were analyzed using 
the FOCAS image software (Valdes, 1982) which extracted both the positions and isophotal 
moments required for the mass inversion. These were then analyzed using the 
method outlined previously, 
the results of which are now discussed.

\section{Results}
\label{sec:results}
%auto-ignore

\subsection{Implementation}
\label{sec:implement}

Before proceeding with the results of the simulation, we briefly
describe the actual implementation of the maximum likelihood method, 
whose formalism
was discussed in \S~3. In what follows, we assume
that the images have been suitably processed with 
extracted galaxy positions and
ellipticities.

An essential element of the maximum likelihood routine is the 
intrinsic ellipticity
distribution of galaxies. Instead of directly 
using the distribution we used to
generate the background fields (Eq. \ref{eq:e_dist}),
we simulated 10 background fields of galaxies
(independent of those used in the actual simulations) and used a fit to 
the observed ellipticity distribution as the intrinsic  ellipticity 
distribution. The best fit obtained was (see also Figure \ref{fig:ellip})
\begin{equation}
P(\epsilon_{S}) \, d\epsilon_S \propto 
\exp(-1.75866(\epsilon_{S}+1.1388)^{2})
\label{eq:e_dist_real}
\end{equation}
with the constant of proportionality determined by the normalization of the 
probability function. Note that this is not directly related to Eq.
\ref{eq:e_dist} since pixelization and other observational effects 
have been convolved into it. We discuss this in more detail later
 in this section.

Each lensed image was analyzed by constructing the 
likelihood function (Eq. \ref{eq:like})
on a $\delta_c$-$\theta_s$ 
($\sigma_v$-$\theta_s$ for SIS analyses) grid assuming the
three (NFW, SIS, KKBP) models in turn. For these three types of analyses,
an expectation value for the mass enclosed in a radius of $1.5$ Mpc was
computed by weighting the mass obtained at every point on the $\delta_c$-
$\theta_s$ grid by its corresponding probability, summing and dividing by a
suitable normalization constant,
\begin{equation}
\langle M_{1500} \rangle = \frac {\sum p(\delta_c,\theta_s) M_{1500}(\delta_c,\theta_s)}
{\sum p(\delta_c,\theta_s)}
\end{equation}
where $M_{1500}$ is the mass at $1.5$ Mpc and $p(\delta_c,\theta_s)$ is the
probability function. We chose a
fixed radius for the mass computation to compare the accuracy of mass
estimation even if the incorrect model was assumed for the analysis. In
the case where the virial mass is desired, an appropriate formalism is
discussed in Appendix B.

The parameter estimation and profile identification used more direct methods.
The location and value of the maximum of the likelihood function
determined the parameters $(\delta_c,\theta_s)_{ML}
$. The profile type was determined by choosing the model with the greatest
probability. The following sections summarize the results thus obtained.

\subsection{Profile Identification}

\begin{table}{\footnotesize
\begin{center}
{{\sc Table I: Profile Identification Fractions}}
\begin{tabular}{rccccc}
\hline\hline
Profile Type & Redshift & & Mass & ($10^{15} M_\odot$) & \\
& & $0.5$ & $0.75$ & $1.0$ & $1.5$ \\
\hline
\textit{SIS}
& 0.2 & 1.0 & 1.0 & 0.8 & 0.9 \\
& 0.5 & 0.9 & 0.7 & 0.9 & 0.8 \\
& 0.8 & 0.8 & 0.8 & 0.8 & 0.8 \\
\hline
\textit{NFW}
& 0.2 & 0.5 & 0.4 & 0.2 & 0.4 \\
& 0.5 & 0.6 & 0.3 & 0.3 & 0.5 \\
& 0.8 & 0.5 & 0.1 & 0.5 & 0.7 \\
\hline
\textit{KKBP}
& 0.2 & 0.5 & 0.6 & 0.7 & 1.0 \\
& 0.5 & 1.0 & 1.0 & 1.0 & 0.7 \\
& 0.8 & 0.8 & 0.9 & 1.0 & 0.7 \\
\hline
\textit{NFW+}
& 0.2 & 0.5 & 0.5 & 0.55 & 0.85 \\
\textit{KKBP}
& 0.5 & 0.85 & 0.8 & 0.9 & 0.85 \\
& 0.8 & 0.7 & 0.65 & 0.95 & 0.95 \\
\hline
&&0.5 & 0.75 & 1.0 & 1.5 \\
\hline
\end{tabular}
\end{center}
\caption {
The  fraction of dark matter profiles identified
correctly, at the specified redshifts and masses. The NFW+KKBP category is the 
fraction of clusters correctly identified as not being SIS.}
\label{tab:ids}}
\end{table}

\begin{figure}
\psfig{figure=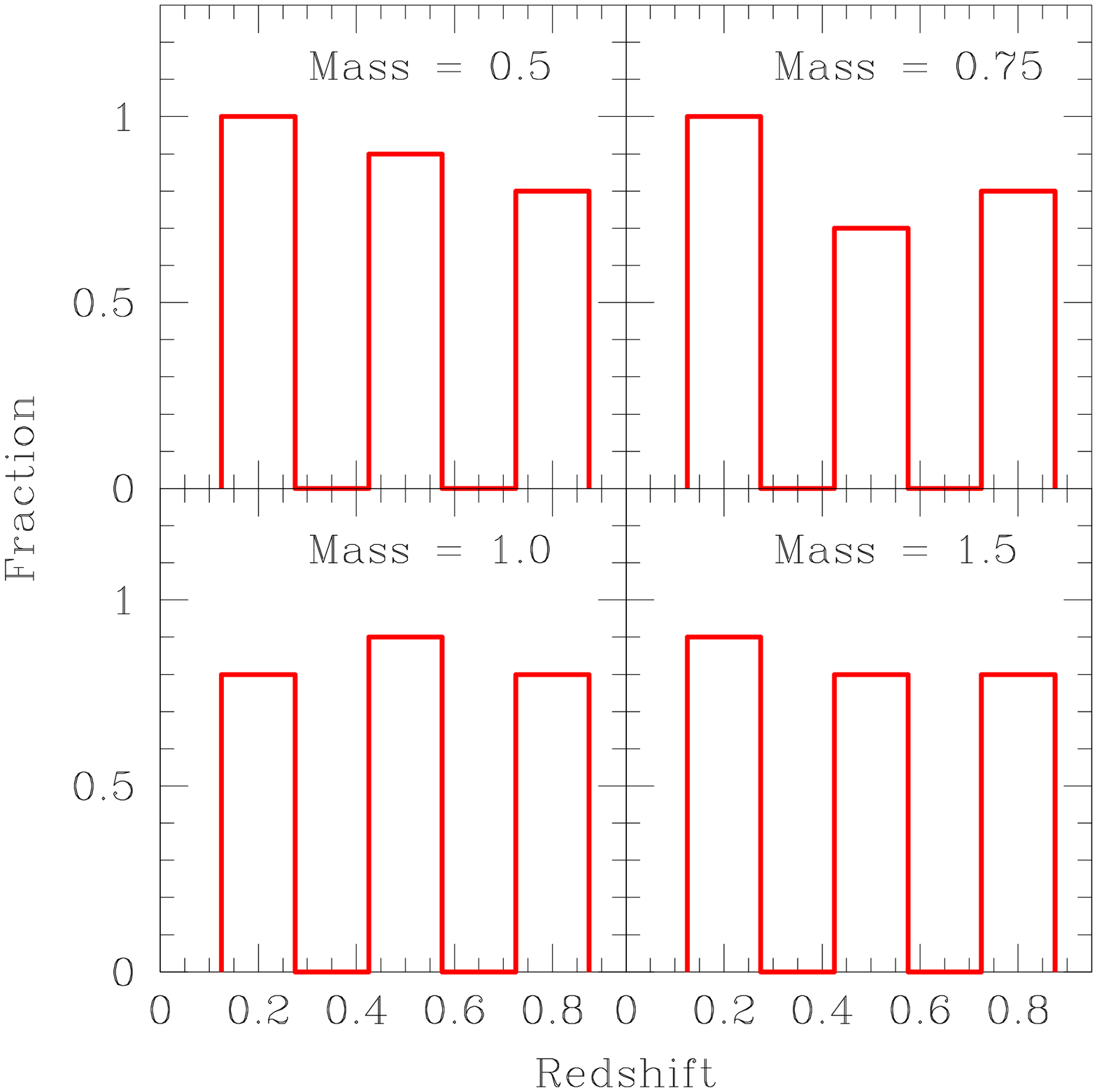,height=6truecm,width=8truecm}
\caption{%
The fraction of SIS clusters correctly identified as such. The masses 
are all $\times 10^{15} M_{\odot}$. 
}
\label{fig:id_sis}
\end{figure}

\begin{figure}
\psfig{figure=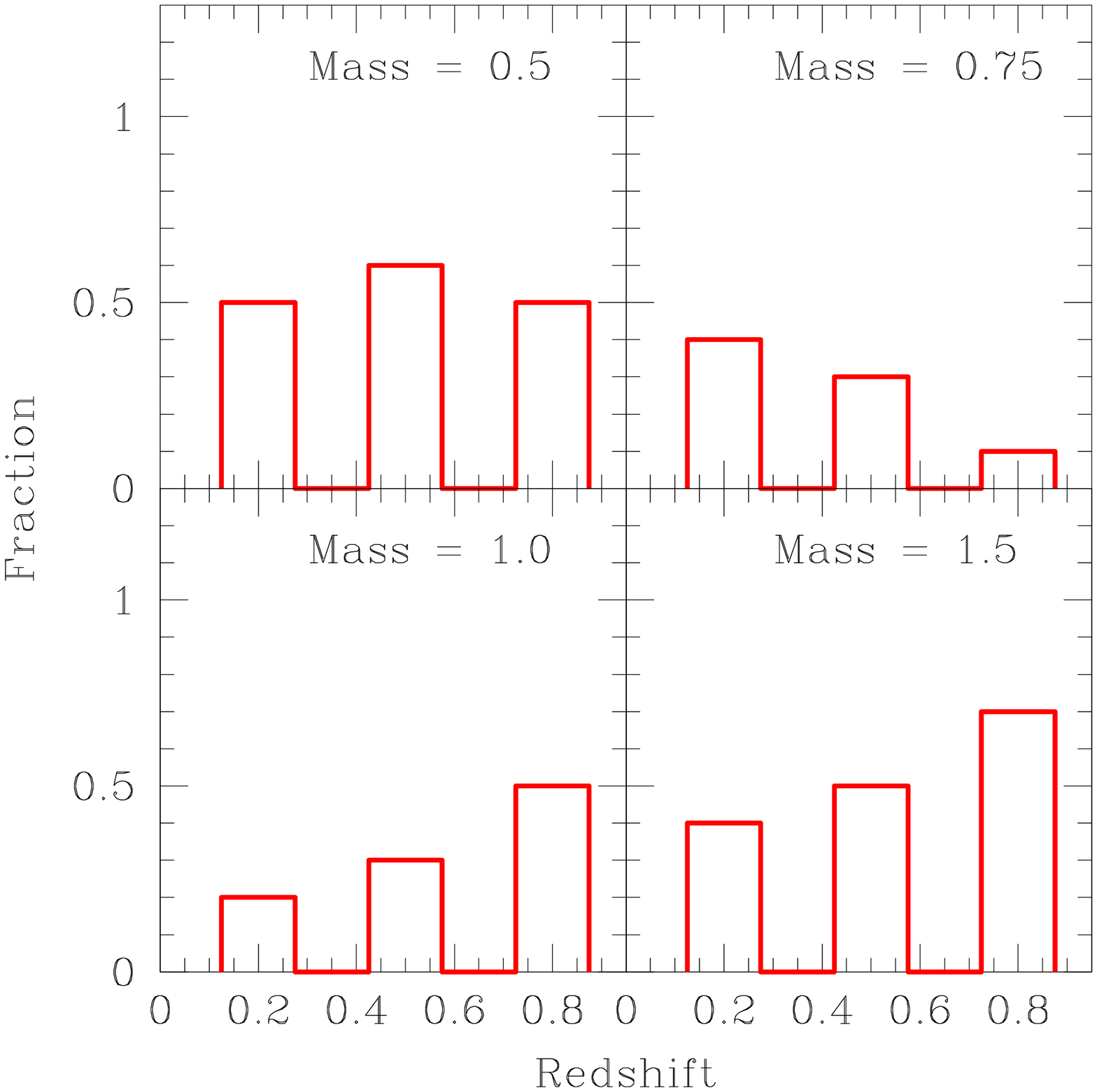,height=6truecm,width=8truecm}
\caption{%
The fraction of NFW clusters correctly identified as such. The masses 
are all $\times 10^{15} M_{\odot}$.
}
\label{fig:id_nfw}
\end{figure}

\begin{figure}
\psfig{figure=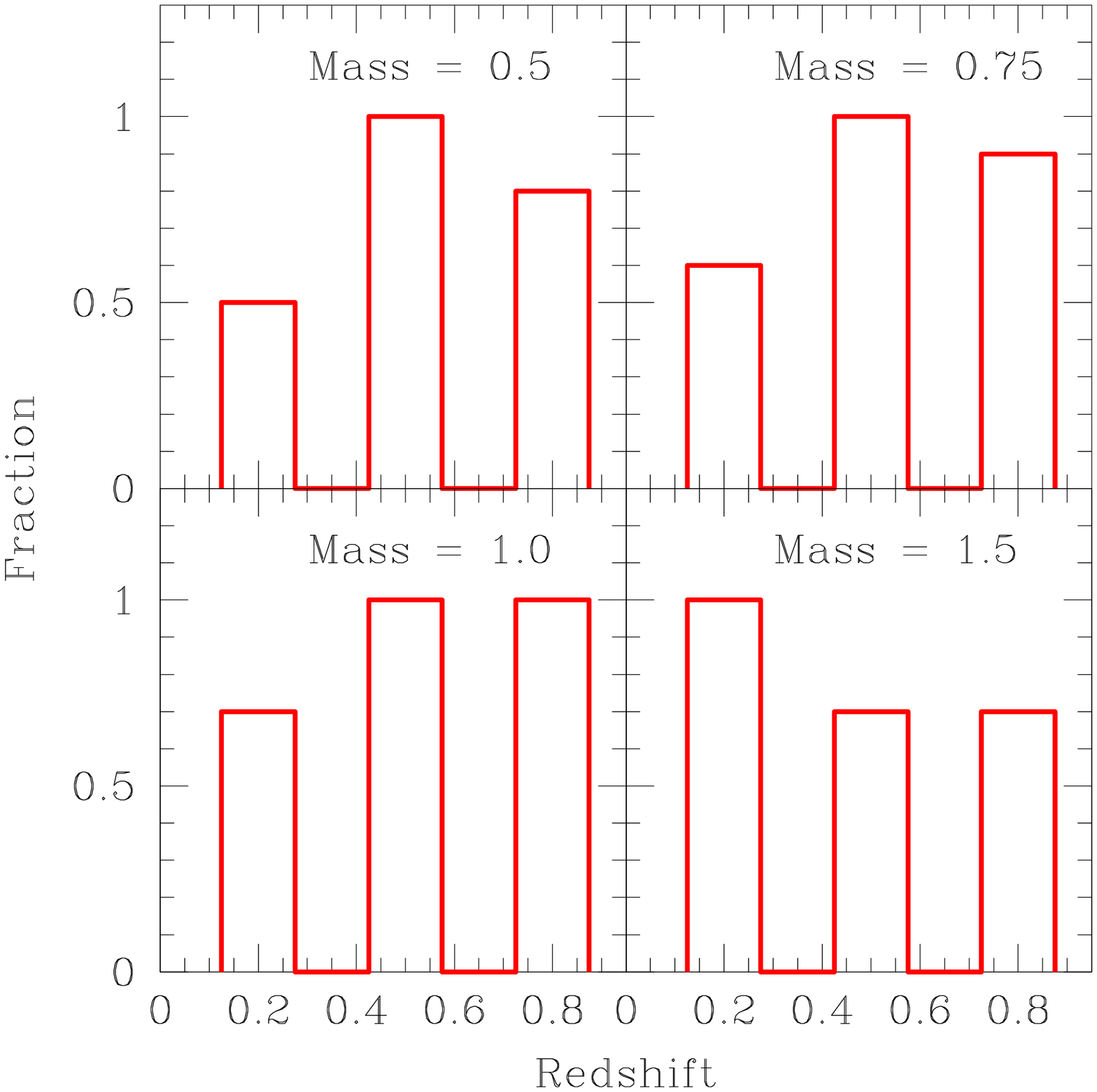,height=6truecm,width=8truecm}
\caption{%
The fraction of KKBP clusters correctly identified as such. The masses 
are all $\times 10^{15} M_{\odot}$.
}
\label{fig:id_kkbp}
\end{figure}

\begin{figure}
\psfig{figure=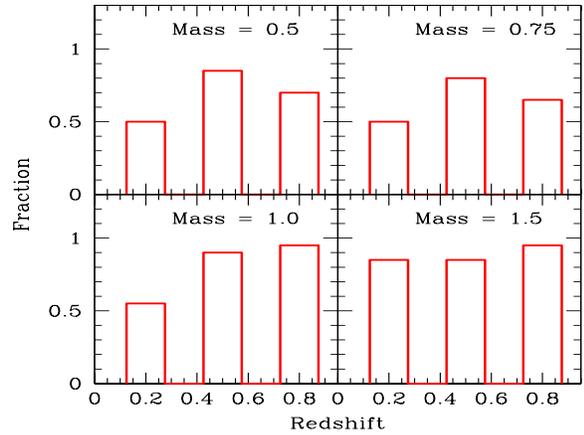,height=6truecm,width=8truecm}
\caption{%
The fraction of clusters correctly identified as not being SIS. The masses 
are all $\times 10^{15} M_{\odot}$.
}
\label{fig:id_nfwkkbp}
\end{figure}

The profile identification results are summarized in Table
\ref{tab:ids} and Figures~\ref{fig:id_sis} to 
\ref{fig:id_nfwkkbp}. We found it
instructive to consider the identification of all three profiles
separately as well as the ability to distinguish between the SIS
and the NFW/KKBP profiles jointly. This was motivated by the
similarities between the NFW and KKBP profiles, which differ only
in the inner core, an area where weak lensing is not 
very sensitive.

As is apparent from the data, our method appears to 
correctly distinguish SIS clusters from the NFW/KKBP joint
profile; the correctly identified fraction being approximately
$80\%$ for clusters at $z=0.8$. 
At low redshifts, both NFW and
KKBP clusters are misidentified as SIS while at higher 
redshifts, NFW clusters tend to be misidentified as 
KKBP.
This implies that our method is, as
expected, unable to distinguish between the NFW and KKBP
profiles, especially at high redshifts.
This can be traced to the fact that the two profiles
only differ in their innermost regions, evident from 
Figure~\ref{fig:compdens}. However, weak lensing is not sensitive 
to these regions of
the cluster, and so to distinguish between such clusters, we must
fall back to strong lensing results. We discuss this further in
\S~6. For clusters at low redshifts, the lensing data does not
extend to a large enough radius to be sensitive to the change in slope
of the NFW and KKBP profiles. Accordingly, a greater number are misidentified
as SIS.

\subsection{Mass Estimation - I}

\begin{table}{\footnotesize
\begin{center}
{\sc Table II: Mass Estimation - I}
\begin{tabular}{rcccc}
\hline\hline
Profile Type & $M_{1500}$ ($10^{15} M_\odot$) & & Redshift & \\
& & 0.2 & 0.5 & 0.8 \\
\hline
\textit{SIS}
& 0.575 & 0.549 & 0.550 & 0.545 \\
& 0.754 & 0.715 & 0.727 & 0.731 \\
& 0.914 & 0.901 & 0.875 & 0.873 \\
& 1.197 & 1.186 & 1.191 & 1.169 \\
\hline
\textit{NFW}
& 0.550 & 0.515 & 0.511 & 0.510 \\
& 0.756 & 0.710 & 0.724 & 0.703 \\
& 0.951 & 0.906 & 0.882 & 0.922 \\
& 1.318 & 1.298 & 1.291 & 1.223 \\
\hline
\textit{KKBP}
& 0.546 & 0.512 & 0.503 & 0.502 \\
& 0.675 & 0.646 & 0.628 & 0.625 \\
& 0.800 & 0.757 & 0.741 & 0.742 \\
& 1.037 & 0.990 & 1.009 & 0.957 \\
\hline
&& 0.2 & 0.5 & 0.8 \\
\hline
\end{tabular}
\end{center}
\caption{
The mean estimated values for $M_{1500}$ if we assume
that the profile was correctly identified.}
\label{tab:mass_actual}}
\end{table}

\begin{figure}
\psfig{figure=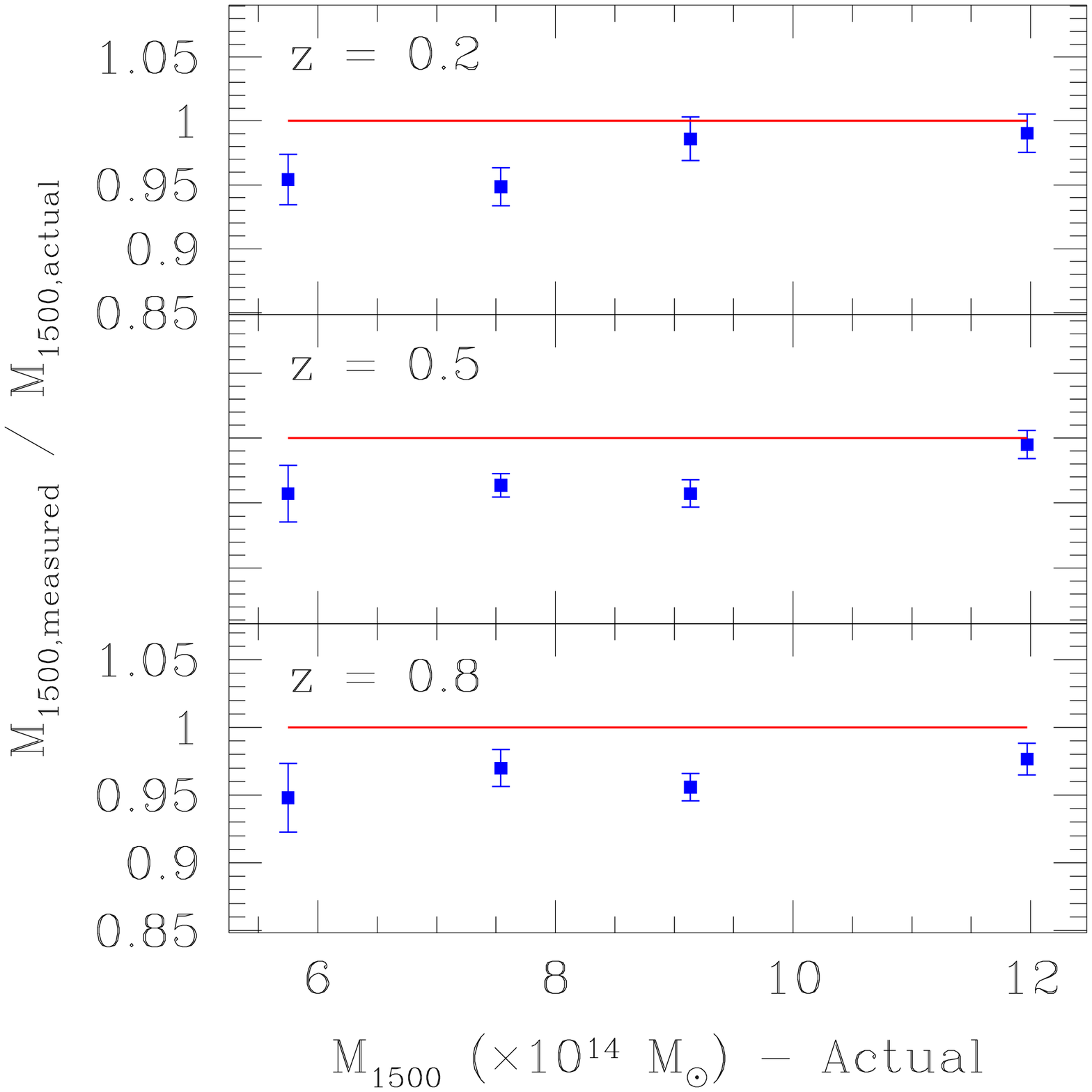,height=6truecm,width=8truecm}
\caption{%
The ratio of the estimated mean mass to the 
true mass of the 10 simulated SIS clusters
against the true masses. The errorbars are the
standard deviation scaled by $1/ \sqrt{10}$. The solid line is y = 1,
depicting equality between the actual and measured masses.
}
\label{fig:mass_act_sis}
\end{figure}

\begin{figure}
\psfig{figure=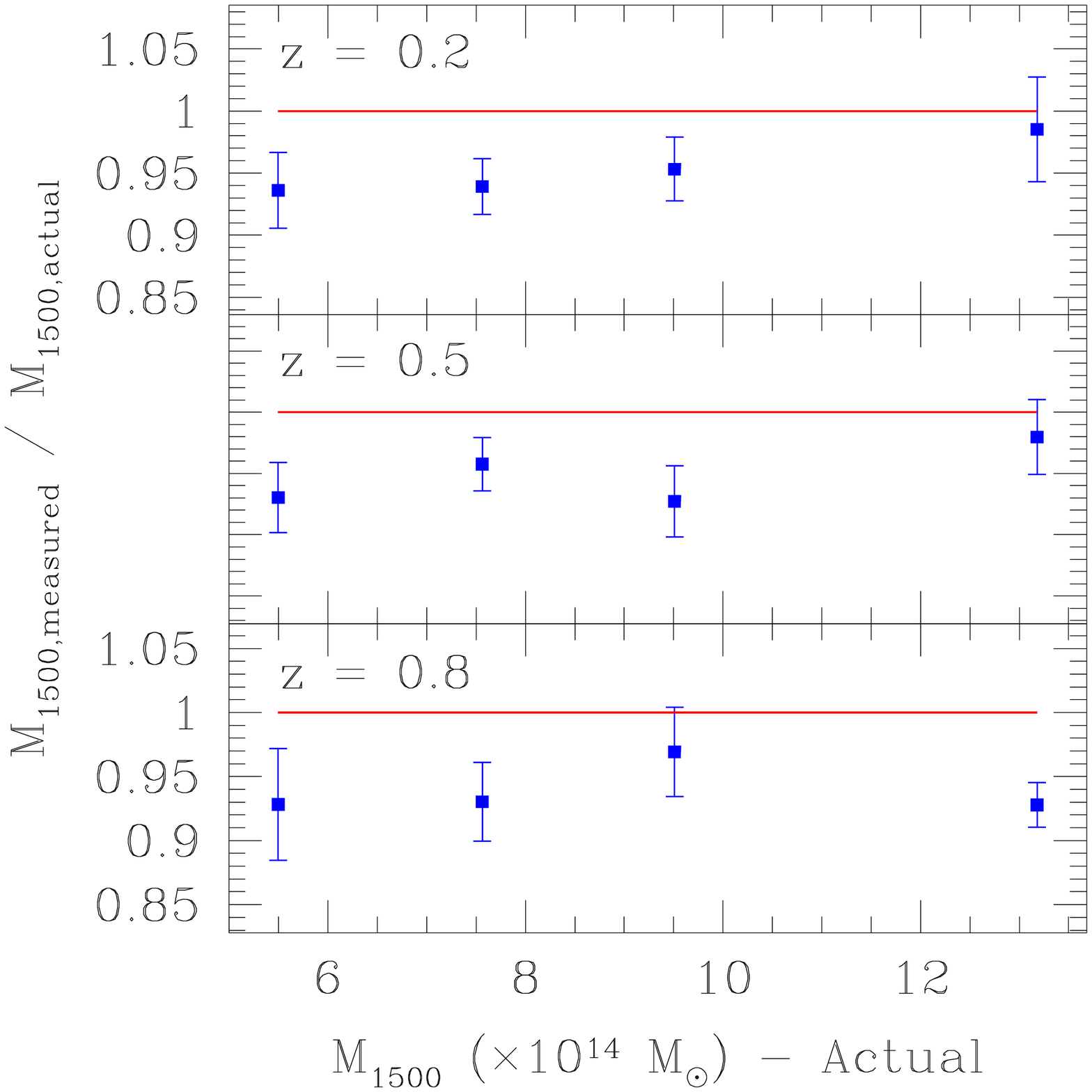,height=6truecm,width=8truecm}
\caption{%
The ratio of the estimated mean mass to the
true mass of the 10 simulated NFW clusters
against the true masses. The errorbars are the
standard deviation scaled by $1/ \sqrt{10}$. The solid line is y = 1,
depicting equality between the actual and measured masses.
}
\label{fig:mass_act_nfw}
\end{figure}

\begin{figure}
\psfig{figure=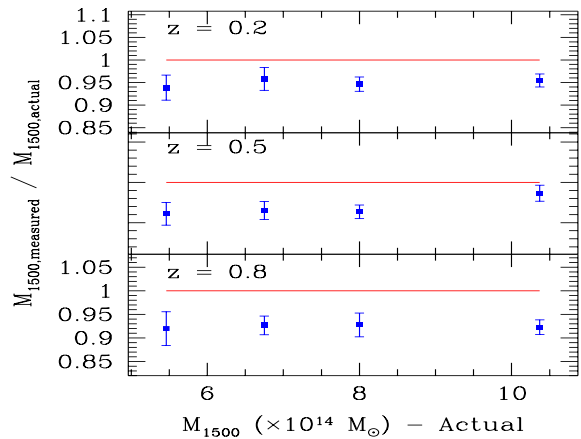,height=6truecm,width=8truecm}
\caption{%
The ratio of the estimated mean mass to the 
true mass of the 10 simulated KKBP clusters,
against the true masses. The errorbars are the
standard deviation scaled by $1/ \sqrt{10}$. The solid line is y = 1,
depicting equality between the actual and measured masses.
}
\label{fig:mass_act_kkbp}
\end{figure}

The mass estimates for each of the clusters are summarized in
Table~\ref{tab:mass_actual} and Figures~\ref{fig:mass_act_sis}, 
\ref{fig:mass_act_nfw} and \ref{fig:mass_act_kkbp}. 
As mentioned earlier, these
estimates correspond to the mass of the cluster at $1.5$ Mpc $\equiv
M_{1500}$. Also, we assumed that the cluster profile was
identified correctly; mass estimates for incorrectly identified
clusters are discussed in the following section.

Table~\ref{tab:mass_actual} shows that the cluster masses are correct
to within a 
$10\%$ error at $M_{1500} \approx 0.5 M_{\odot}$ that
improves to $\leq 5\%$ at $M_{1500} \approx 1.5 M_{\odot}$. The
dependence of these estimates on redshift appears to be weak
and there is no clear trend. This can possibly be attributed to the fact 
that although the physical extent of the images increases with
increasing redshift, the signal falls off with increasing radial distance
from the center of the cluster. This loss of signal with increasing redshift
is also evident from an examination of the diagonal elements of Table \ref{tab:bias}.  

An interesting feature of the data presented is the systematic
$5\%$ bias in the mass estimates. Further
numerical experiments pointed to the input
ellipticity distribution as the cause of this bias. This dependence of the
bias on the ellipticity distribution can be explained as follows: imagine we had 
assumed an ellipticity distribution of a $\delta$ function at zero ellipticity. 
In such a case, all mass estimates would be biased above since random elliptical 
galaxies would be erroneously interpreted as a weak lensing signal, indicating
a mass distribution where none exists. Similarly, if we had assumed an intrinsic 
ellipticity distribution that had a greater fraction of elliptical galaxies than 
actually existed, such a distribution would result in an underestimate of the 
lensing and therefore the mass distribution. 

It would appear, at first sight, that in our ``experiments'' we
could avoid this by simply including our input ellipticity
distribution, i.e. use Eq. \ref{eq:e_dist} instead of Eq. 
\ref{eq:e_dist_real}.
However, due to the finite size of galaxies and
pixelization effects, the input ellipticity distribution and the
observed ellipticity distribution are not the same. Our results
correspond to estimates made with fits to the
observed distribution, a choice further justified by the fact
that any real observations must rely on such observed
distributions since the true intrinsic distribution
is unknown. This choice was borne out by numerical 
experiments that showed that estimates based on our Gaussian fit were
more accurate than those from Eq. \ref{eq:e_dist}.

\subsection{Mass Estimation - II}

\begin{table}{ \footnotesize
\begin{center}
{\sc Table III: Mass Estimation - II - SIS clusters}
\begin{tabular}{rccc}
\hline\hline
Redshift & $M_{1500} (10^{15} M_\odot)$ & Profiles & \\
& & \textit{NFW} & \textit{KKBP} \\
\hline
z = 0.2
 & 0.575 & 0.300 & 0.224 \\
 & 0.754 & 0.425 & 0.311 \\
 & 0.914 & 0.541 & 0.404 \\
 & 1.197 & 0.755 & 0.590 \\
\hline
z = 0.5
 & 0.575 & 0.382 & 0.285 \\
 & 0.754 & 0.522 & 0.405 \\
 & 0.914 & 0.640 & 0.525 \\
 & 1.197 & 0.906 & 0.780 \\
\hline
z = 0.8
& 0.575 & 0.378 & 0.279 \\
 & 0.754 & 0.552 & 0.410 \\
 & 0.914 & 0.669 & 0.531 \\
 & 1.197 & 0.889 & 0.758 \\
\hline
& & \textit{NFW} & \textit{KKBP} \\
\hline
\end{tabular}
\end{center}
\caption{
The mean estimated masses for the SIS clusters
when analyzed assuming an incorrect profile.}
\label{tab:mass_wrong_sis}}
\end{table}

\begin{table}{ \footnotesize
\begin{center}
{\sc Table IV:Mass Estimation - II - NFW clusters}
\begin{tabular}{rccc}
\hline\hline
Redshift & $M_{1500} (10^{15} M_\odot)$ & Profiles & \\
& & \textit{SIS} & \textit{KKBP} \\
\hline
z = 0.2
 & 0.550 & 0.717 & 0.331 \\
 & 0.756 & 1.012 & 0.488 \\
 & 0.951 & 1.253 & 0.636 \\
 & 1.318 & 1.722 & 0.991 \\
\hline
z = 0.5
& 0.550 & 0.688 & 0.378 \\
 & 0.756 & 0.912 & 0.579 \\
 & 0.951 & 1.120 & 0.742 \\
 & 1.318 & 1.607 & 1.110 \\
\hline
z = 0.8
& 0.550 & 0.673 & 0.394 \\
 & 0.756 & 0.872 & 0.552 \\
 & 0.951 & 1.111 & 0.755 \\
 & 1.318 & 1.565 & 1.096 \\
\hline
& & \textit{SIS} & \textit{KKBP} \\
\hline
\end{tabular}
\end{center}
\caption{
The mean estimated masses for the NFW clusters
when analyzed assuming an incorrect profile.}
\label{tab:mass_wrong_nfw}
}
\end{table}

\begin{table}{ \footnotesize

\begin{center}
{\sc Table V:Mass Estimation - II - KKBP clusters}
\begin{tabular}{rccc}
\hline\hline
Redshift & $M_{1500} (10^{15} M_\odot)$ & Profiles & \\
& & \textit{SIS} & \textit{NFW} \\
\hline
z = 0.2
& 0.546 & 0.943 & 0.734 \\
 & 0.675 & 1.166 & 0.943 \\
 & 0.800 & 1.364 & 1.135 \\
 & 1.037 & 1.738 & 1.347 \\
\hline
z = 0.5
& 0.546 & 0.797 & 0.611 \\
 & 0.675 & 0.966 & 0.760 \\
 & 0.800 & 1.124 & 0.896 \\
 & 1.037 & 1.433 & 1.119 \\
\hline
z = 0.8
& 0.546 & 0.766 & 0.582 \\
 & 0.675 & 0.942 & 0.735 \\
 & 0.800 & 1.087 & 0.860 \\
 & 1.037 & 1.331 & 1.080 \\
\hline
& & \textit{SIS} & \textit{NFW} \\
\hline
\end{tabular}
\end{center}
\caption{
The mean estimated masses for the KKBP clusters
when analyzed assuming an incorrect profile.}
\label{tab:mass_wrong_kkbp}}
\end{table}

\begin{table} { \footnotesize

\begin{center}
{\sc Table VI: Average percentage bias}
\begin{tabular}{rcccc}
\hline\hline
Redshift & Actual & & Bias & \\
\hline
& & \textit{SIS} & \textit{NFW} & \textit{KKBP} \\
\hline
z = 0.2 
& \textit{SIS} & -2.2 & -40.5 & -55.2 \\
& \textit{NFW} & 32.0 & -4.2 & -31.6 \\
& \textit{KKBP} & 70.3 & 38.3 & -4.9 \\
\hline
z = 0.5 
& \textit{SIS} & -3.1 & -28.7 & -41.6 \\
& \textit{NFW} & 19.5 & -5.2 & -20.8 \\
& \textit{KKBP} & 40.6 & 11.1 & -6.1 \\
\hline
z = 0.8 
& \textit{SIS} & -3.6 & -26.5 & -41.5 \\
& \textit{NFW} & 16.9 & -5.1 & -19.6 \\
& \textit{KKBP} & 34.9 & 7.0 & -7.4 \\
\hline
& & \textit{SIS} & \textit{NFW} & \textit{KKBP} \\
\hline
\end{tabular}
\end{center}
\caption{
The observed percentage bias in $M_{1500}$ estimation for 
various assumed profiles.}
\label{tab:bias}}
\end{table}

\begin{figure}
\psfig{figure=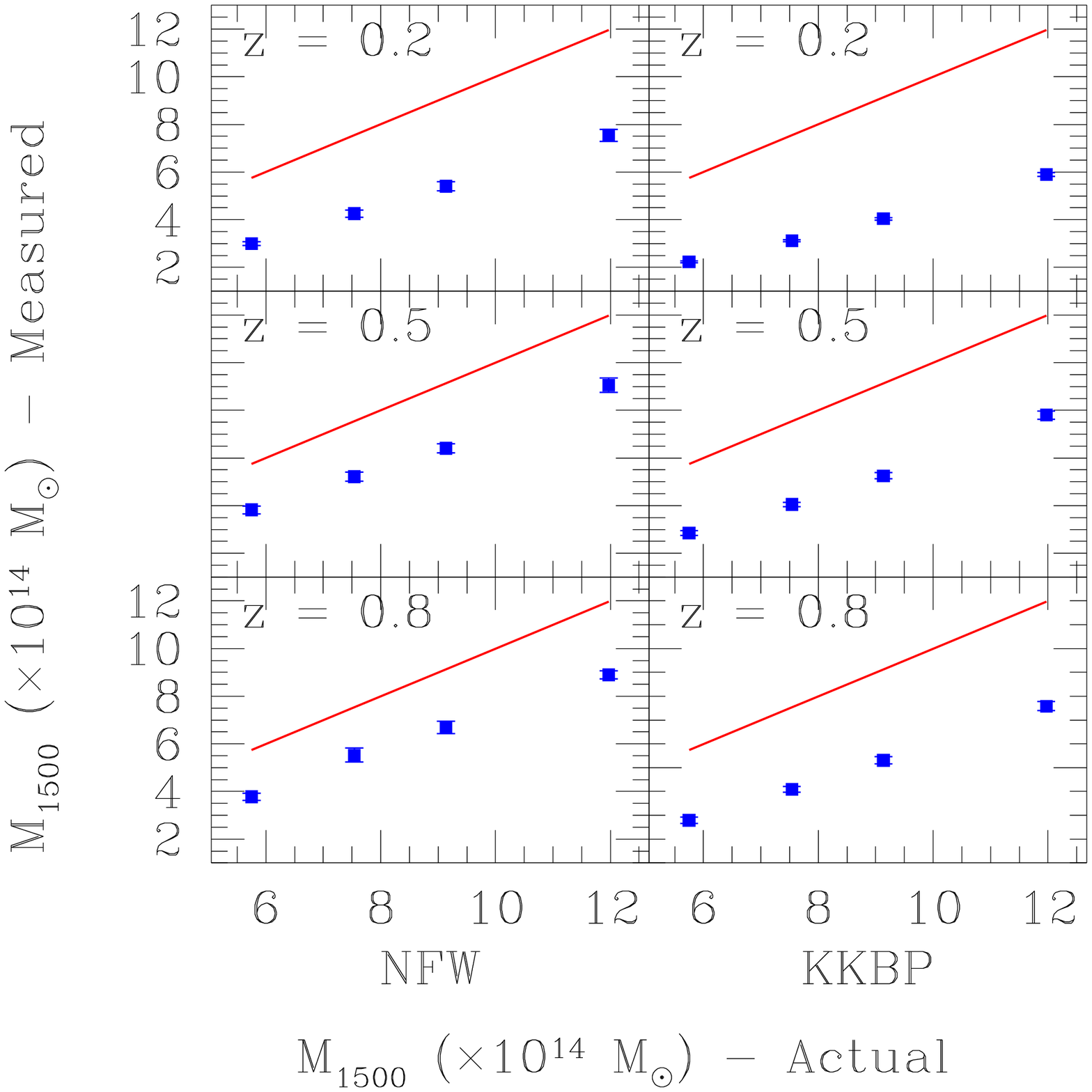,height=6truecm,width=8truecm}
\caption{%
The estimated mean masses of the 10 simulated SIS clusters against
the actual masses, if analyzed
assuming an NFW or KKBP profile. The errorbars are the
standard deviation scaled by $1/ \sqrt{10}$. The solid line has slope
$1$ and zero intercept, depicting equality between the actual and measured masses.
}
\label{fig:mass_wrong_sis_fig}
\end{figure}

\begin{figure}
\psfig{figure=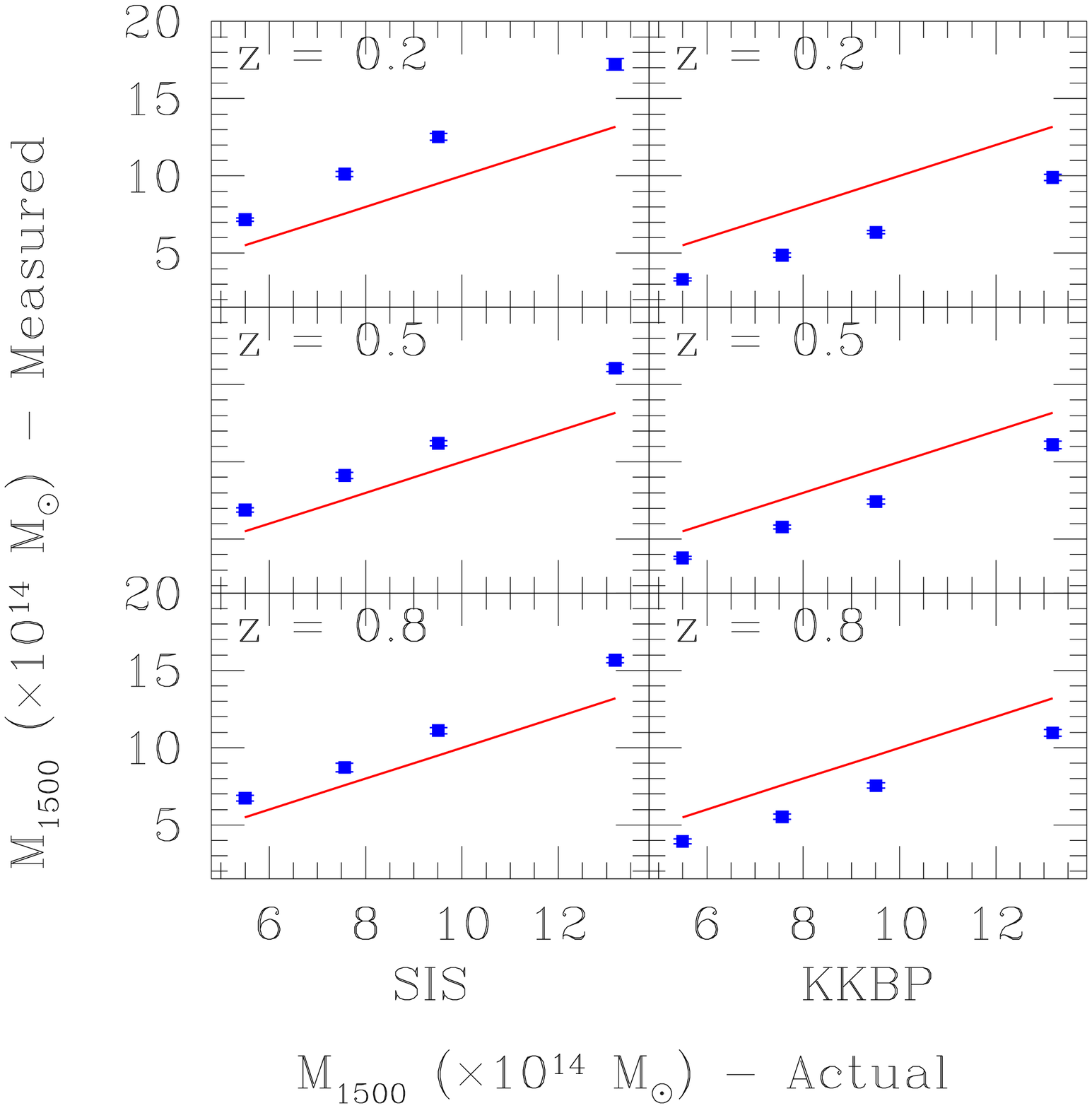,height=6truecm,width=8truecm}
\caption{%
The estimated mean masses of the 10 simulated NFW clusters against
the true masses, if analzyed
assuming an SIS or KKBP profile. The errorbars are the
standard deviation scaled by $1/ \sqrt{10}$. The solid line has slope
$1$ and zero intercept, depicting equality between the actual and measured masses.
}
\label{fig:mass_wrong_nfw_fig}
\end{figure}

\begin{figure}
\psfig{figure=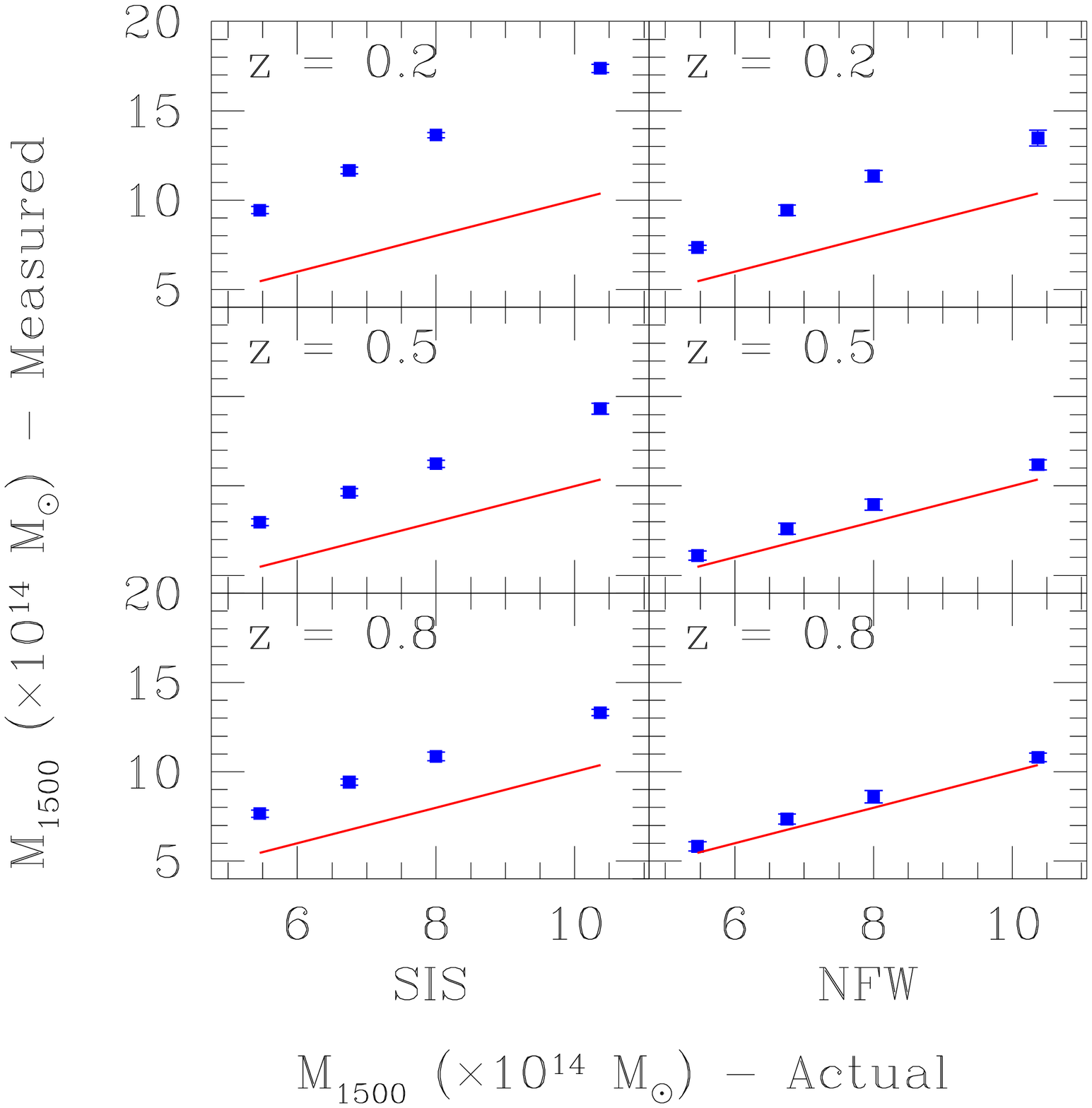,height=6truecm,width=8truecm}
\caption{%
The estimated mean masses of the 10 simulated KKBP clusters against
the true masses, if analzyed
assuming an SIS or NFW profile. The errorbars are the
standard deviation scaled by $1/ \sqrt{10}$. The solid line has slope
$1$ and zero intercept, depicting equality between the actual and measured masses.
}
\label{fig:mass_wrong_kkbp_fig}
\end{figure}

The mass estimations discussed   in the previous section assumed  that
the profile  of the cluster was correctly  identified. It is
equally  important to   quantify the  bias   in the  mass  estimation
if the incorrect profile is used,
especially  if     we  consider  using    these  cluster   masses  for
cosmology. Tables~\ref{tab:mass_wrong_sis},\ref{tab:mass_wrong_nfw}  and
\ref{tab:mass_wrong_kkbp}    and       Figures~\ref{fig:mass_wrong_sis_fig},
\ref{fig:mass_wrong_nfw_fig}, and \ref{fig:mass_wrong_kkbp_fig} 
summarize the mass
estimation in our  simulations when  an incorrect profile  is
assumed. We also list the average  bias as a function of redshift
in Table~\ref{tab:bias}.  A clear trend is  the reduction in the bias with
increasing redshift; for instance the  bias in KKBP clusters  analysed
as SIS clusters   drops from 70\% at $z=0.2$   to 35\% at  $z=0.8$,  a
reduction by  a factor of two.  This  is easily explained by  the fact
that the degree   of extrapolation decreases with increasing  redshift
since more of the cluster is visible and  therefore the mass is better
determined. Note also that the bias in an NFW/KKBP
family cluster analysed with  an SIS profile is  of the order of  30\%
even at a redshift of 0.8,  while the bias drops  to 15\% if
the   assumed profile  was   incorrect  but   part  of  the   NFW/KKBP
family. There is therefore a clear  advantage to having some degree of
profile determination before extrapolating. 

\subsection{Parameter Estimation}

\begin{table} {\footnotesize

\begin{center}
{\sc Table VII:$\delta_{c}$ / $\sigma_{v}$ Estimation}
\begin{tabular}{rcccc}
\hline\hline
Profile Type & $\delta_{c}$ ($\times 10^{4}$) /  & & Redshift & \\
& $\sigma_{v} $ ($\times 10^3$ km/s)& 0.2 & 0.5 & 0.8 \\
\hline
\textit{SIS}
& 0.913 & 0.886 & 0.887 & 0.887 \\
& 1.045 & 1.016 & 1.023 & 1.029 \\
& 1.150 & 1.140 & 1.116 & 1.128 \\
& 1.317 & 1.310 & 1.319 & 1.302 \\
\hline
\textit{NFW}
& 1.57 & 1.59 & 1.68 & 1.72 \\
& 2.15 & 2.28 & 1.92 & 2.12 \\
& 2.70 & 2.80 & 3.19 & 2.45 \\
& 3.75 & 3.68 & 3.63 & 4.35 \\
\hline
\textit{KKBP}
& 2.29 & 2.34 & 2.41 & 2.41 \\
& 2.84 & 2.76 & 2.93 & 3.10 \\
& 3.36 & 3.44 & 3.50 & 3.59 \\
& 4.36 & 4.28 & 4.03 & 4.60 \\
\hline
&& 0.2 & 0.5 & 0.8 \\
\hline
\end{tabular}
\end{center}
\caption{
The mean estimated values for $\delta_{c}$ ($\times 10^4$) for NFW and KKBP
clusters and $\sigma_{v}$ ($\times 10^3$ km/s) for SIS clusters. The profiles are
assumed to be correctly identified.}
\label{tab:delta_c}}
\end{table}

\begin{figure}
\psfig{figure=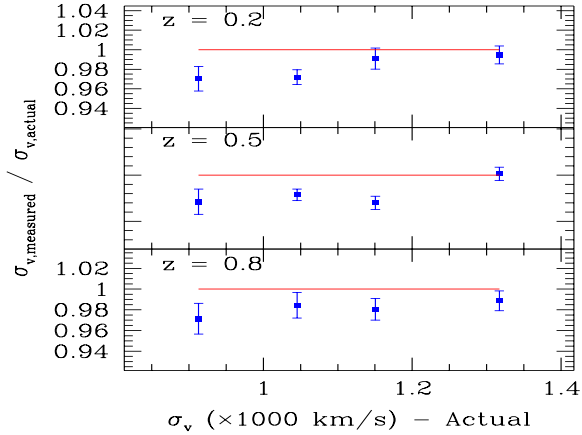,height=6truecm,width=8truecm}
\caption{%
The ratio of the estimated $\sigma_{v}$ to the
actual $\sigma_{v}$ of the 10 simulated SIS clusters
plotted against the true values.
The errorbars are the standard deviation scaled by $1/ \sqrt{10}$. The solid line is
y = 1, depicting equality between actual and measured $\sigma_{v}$.
}
\label{fig:dc_sis}
\end{figure}

\begin{figure}
\psfig{figure=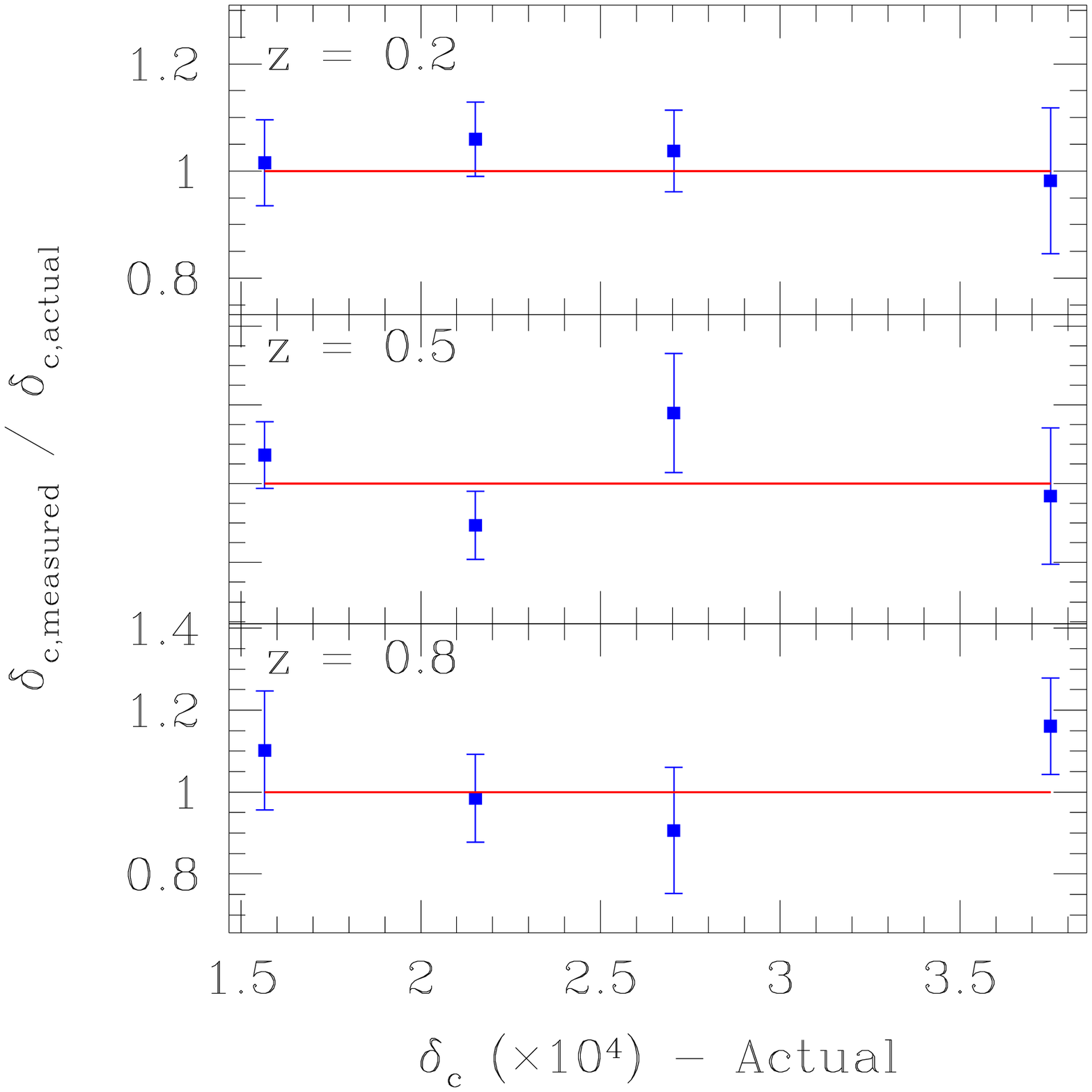,height=6truecm,width=8truecm}
\caption{%
The ratio of the estimated $\delta_{c}$ to the actual 
$\delta_{c}$ of the 10 simulated NFW clusters
plotted against the true values.
The errorbars are the standard deviation scaled by $1/ \sqrt{10}$. The solid line is
y = 1, depicting equality between actual and measured $\delta_{c}$.
}
\label{fig:dc_nfw}
\end{figure}

\begin{figure}
\psfig{figure=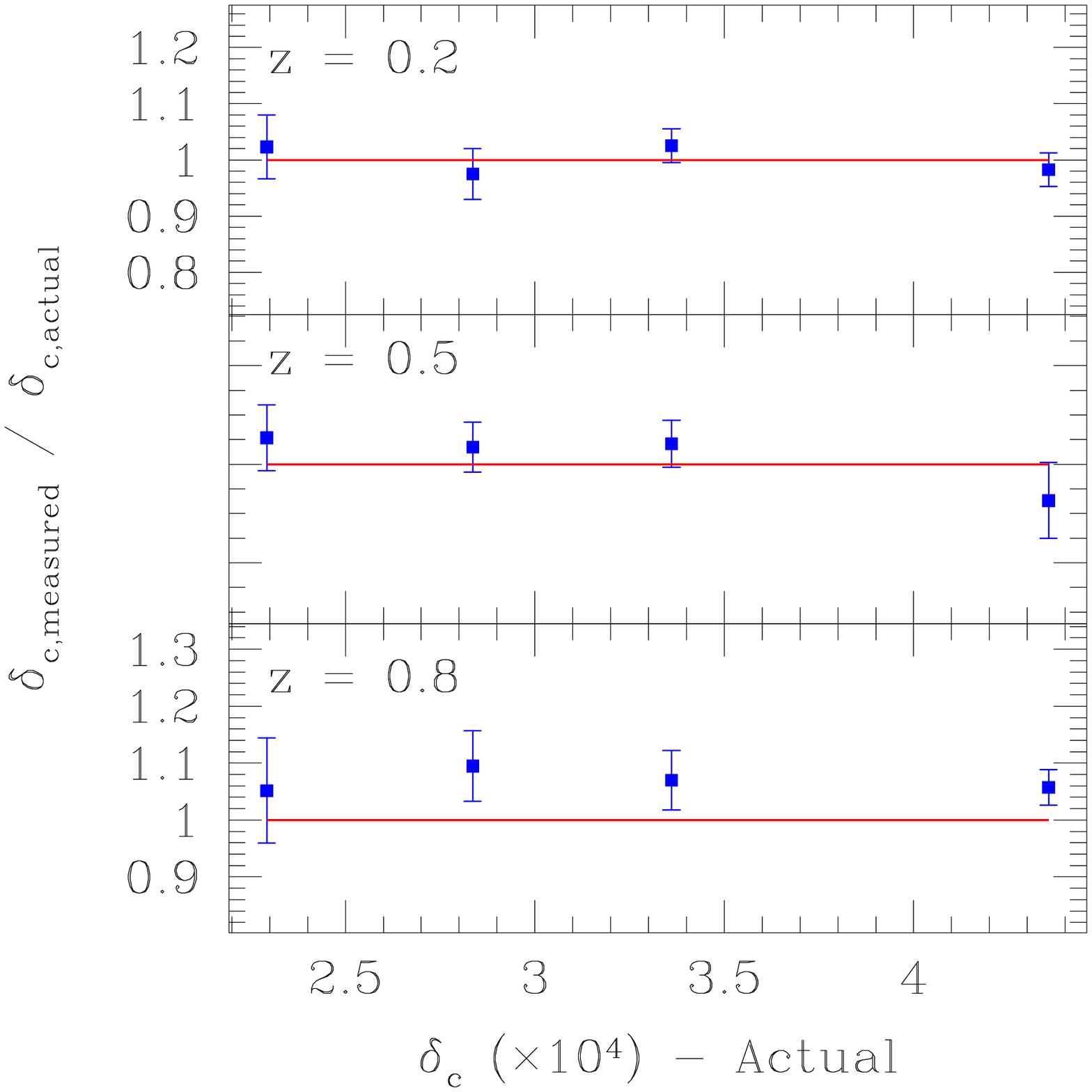,height=6truecm,width=8truecm}
\caption{%
The ratio of the estimated $\delta_{c}$ to the actual
$\delta_{c}$
of the 10 simulated KKBP clusters
plotted against the true values.
The errorbars are the standard deviation scaled by $1/ \sqrt{10}$. The solid line is
y = 1, depicting equality between actual and measured $\delta_{c}$.
}
\label{fig:dc_kkbp}
\end{figure}

\begin{table} {\footnotesize
\begin{center}
{\sc Table VIII:$\theta_{s}$ Estimation}
\begin{tabular}{rcccccc}
\hline\hline
Profile Type & $\theta_{s,true}$ & & Mass & ($10^{15} M_\odot$) & \\
& (arcsec) & 0.5 & 0.75 & 1.0 & 1.5 \\
\hline
\textit{SIS}
& 4.3 & 3.5 & 3.6 & 5.1 & 5.1 \\
& 2.3 & 2.3 & 2.1 & 1.5 & 4.0 \\
& 1.9 & 1.9 & 2.5 & 2.4 & 2.7 \\
\hline
\textit{NFW}
& 86.6 & 85.7 & 83.5 & 86.0 & 97.7 \\
& 46.8 & 45.0 & 50.2 & 45.0 & 54.0 \\
& 38.0 & 37.5 & 38.7 & 44.2 & 36.2 \\
\hline
\textit{KKBP}
& 64.5 & 62.7 & 64.5 & 62.5 & 64.0 \\
& 34.9 & 33.5 & 33.7 & 33.5 & 36.5 \\
& 28.3 & 27.5 & 26.7 & 27.0 & 27.0 \\
\hline
&& 0.5 & 0.75 & 1.0 & 1.5 \\
\hline
\end{tabular}
\end{center}
\caption{
The mean estimated values for $\theta_{s}$ in arcseconds. The profiles are assumed to be
correctly identified.}
\label{tab:theta_s}}
\end{table}

\begin{figure}
\psfig{figure=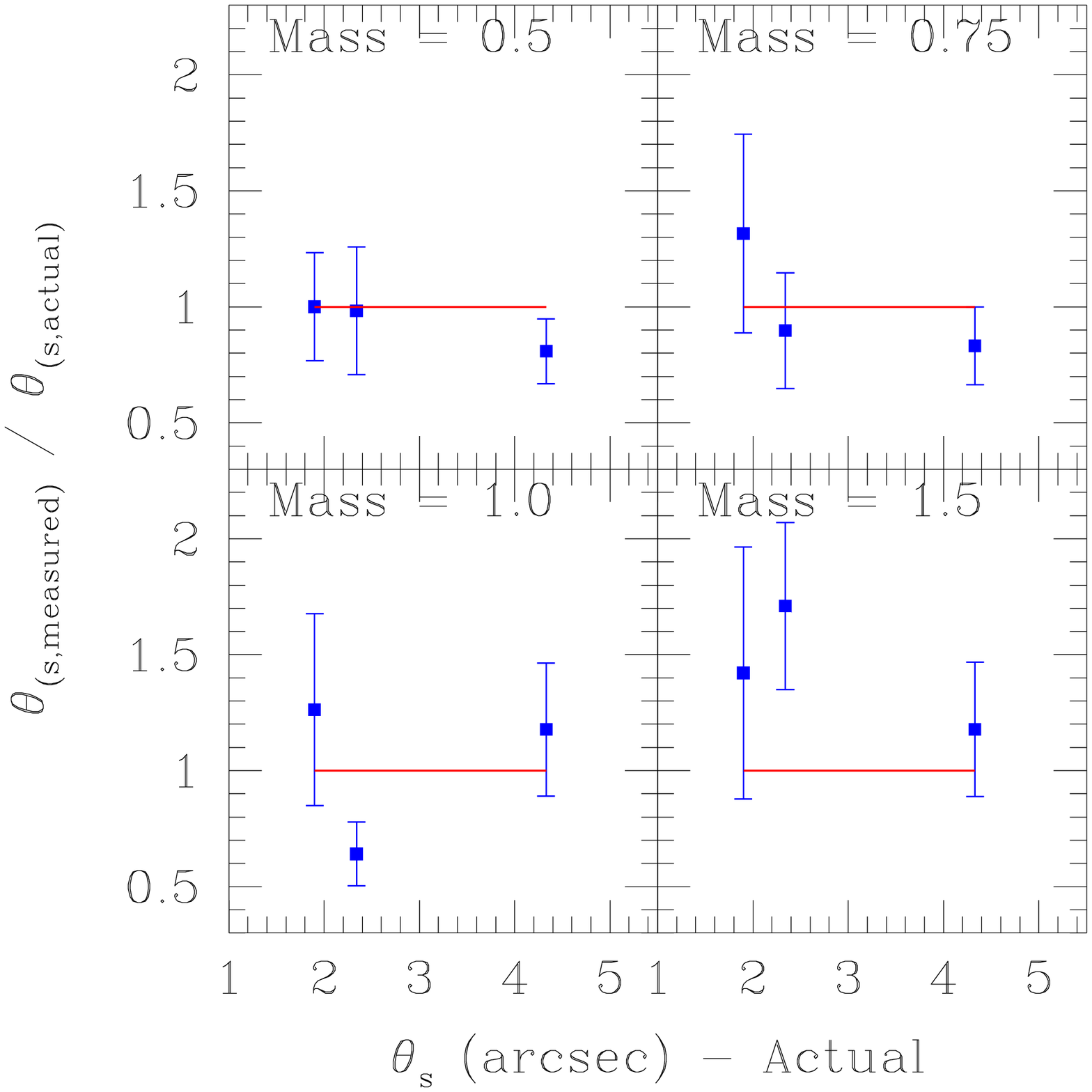,height=6truecm,width=8truecm}
\caption{%
The ratio of the estimated $\theta_{s}$ 
to the actual $\theta_{s}$ of the 10 simulated SIS clusters
plotted against the true values.
The errorbars are the standard deviation scaled by $1/ \sqrt{10}$. The solid line is
y = 1, depicting equality between actual and measured $\theta_{s}$.
}
\label{fig:ts_sis}
\end{figure}

\begin{figure}
\psfig{figure=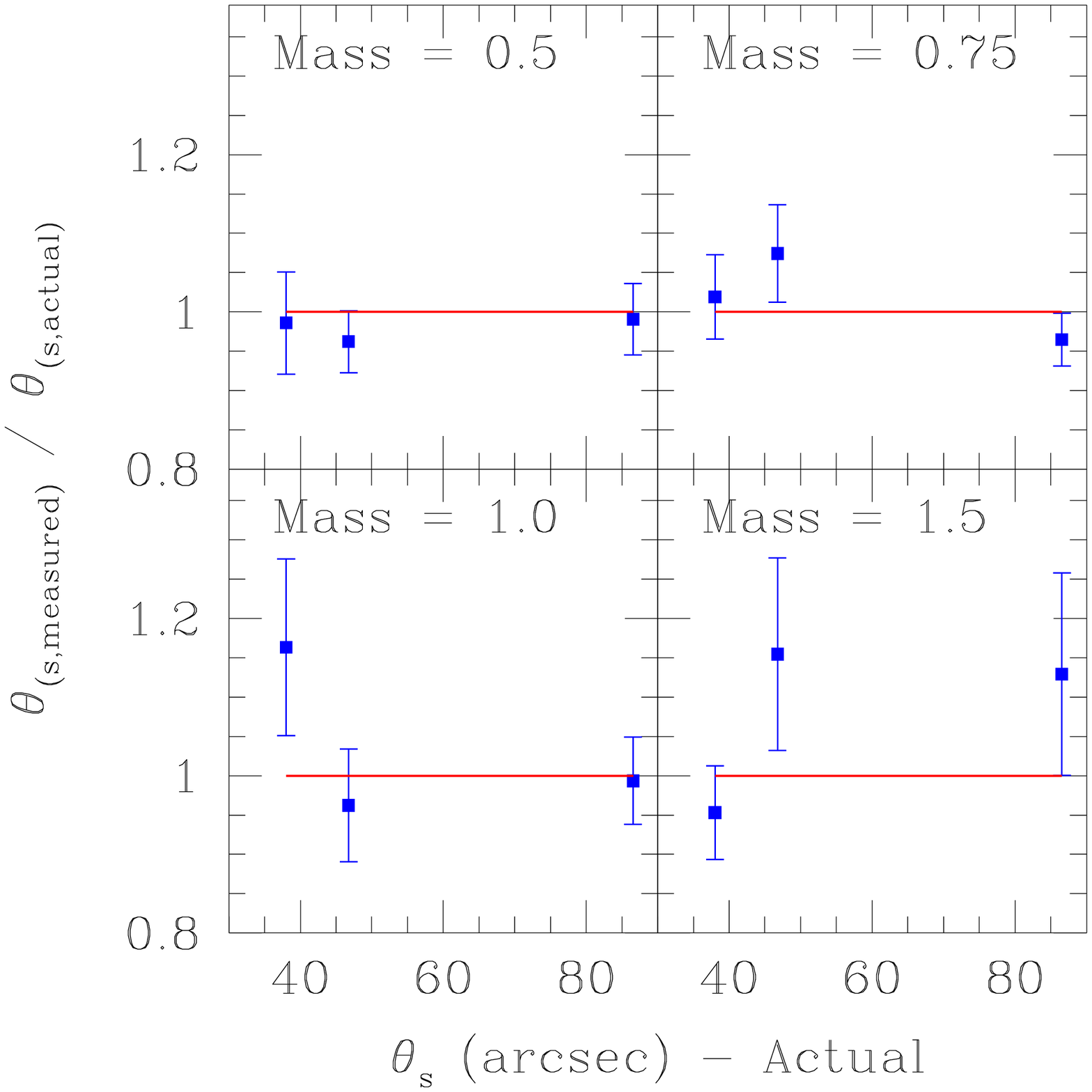,height=6truecm,width=8truecm}
\caption{%
The ratio of the estimated $\theta_{s}$ 
to the actual $\theta_{s}$ of the 10 simulated NFW clusters
plotted against the true values.
The errorbars are the standard deviation scaled by $1/ \sqrt{10}$. The solid line is
y = 1, depicting equality between actual and measured $\theta_{s}$.
}
\label{fig:ts_nfw}
\end{figure}

\begin{figure}
\psfig{figure=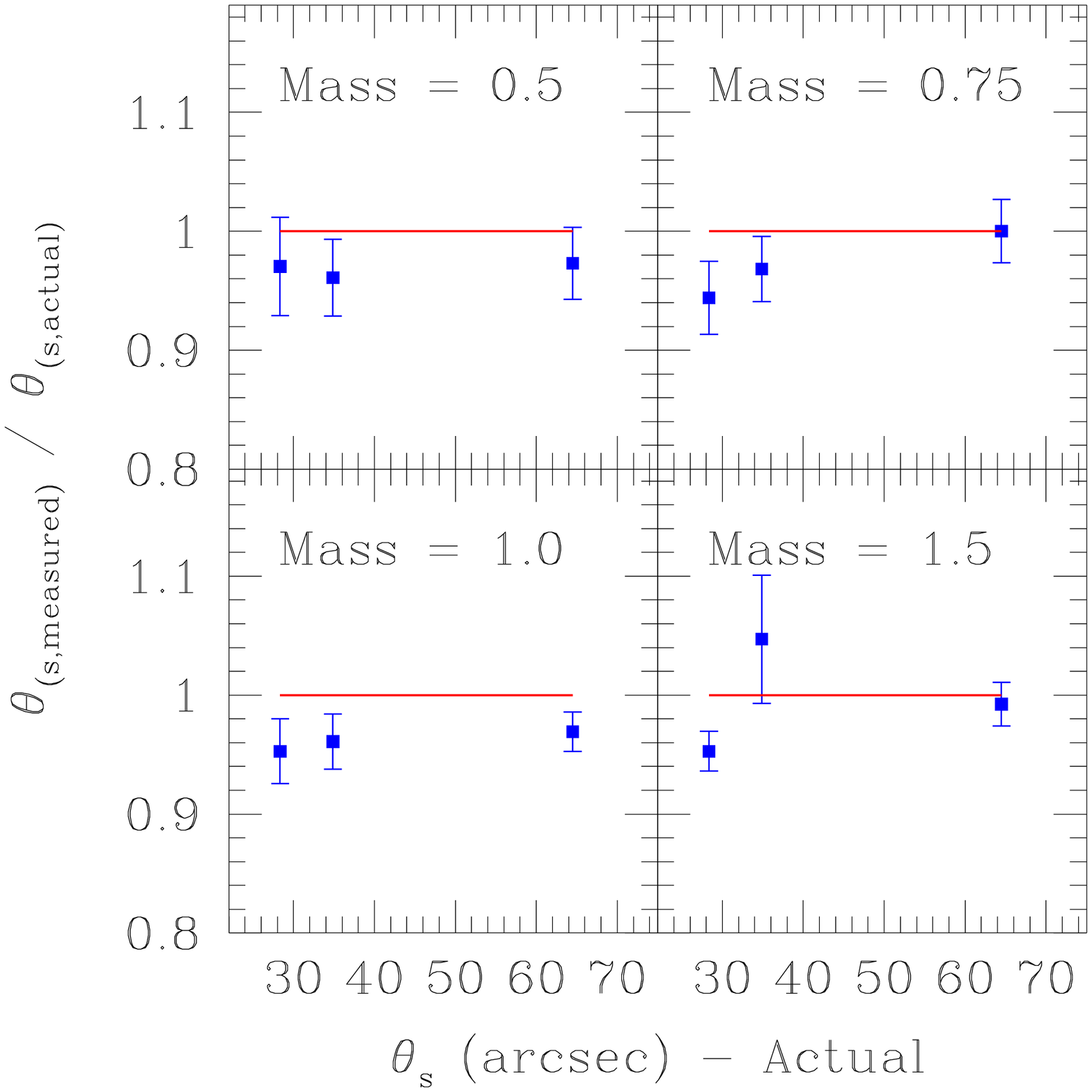,height=6truecm,width=8truecm}
\caption{%
The ratio of the estimated $\theta_{s}$ 
to the actual $\theta_{s}$ of the 10 simulated KKBP clusters
plotted against the true values.
The errorbars are the standard deviation scaled by $1/ \sqrt{10}$. The solid line is
y = 1, depicting equality between actual and measured $\theta_{s}$.
}
\label{fig:ts_kkbp}
\end{figure}

The parameter estimates for  our simulations are summarized in  
Tables~\ref{tab:delta_c}        
and \ref{tab:theta_s}  and Figures~\ref{fig:dc_sis} to
\ref{fig:ts_kkbp}. The most striking feature  about these data is that the
parameters are estimated to within $25\%$ even though
the masses were estimated with a $\leq 10\%$ error. This is explained by
realizing that there is a degeneracy in determining the parameters for a given mass,
increasing the variance in the parameter estimation.
However,  the correlated errors in the parameters
combine to give a smaller cumulative error in the  mass. 

It is also relevant  to note that the  errors in the estimation of the
core    radius for SIS   clusters  are significantly  greater than the
corresponding  errors in the scale  radius for NFW/KKBP clusters. This
is a result of the fact that  the core radius  for the SIS clusters is
$\approx 10$ kpc and  therefore has little  effect on the observed weak
lensing. On the other hand, the scale radius  in the NFW/KKBP clusters
affects   the mass  profile  of the   cluster at significantly greater
scales  and  is therefore  probed  by  weak   lensing. Any attempt  to
robustly estimate the core radius for SIS clusters must rely on strong
lensing.

\section{Discussion}
\label{sec:conclusion}
%auto-ignore

We  have studied the   reliability of mass profile determinations  for
clusters of galaxies using  a parametric  inversion method.  We  found
that the mass profiles of different ``families''  of profiles could be
reliably determined  to  within $80$\%. We appear to able 
to distinguish profiles motivated by N-body CDM simulations
(in our case, NFW and KKBP) from the
canonical $r^{-2}$ SIS profile, making our method a promising approach
to test this prediction. However, we were unable to distinguish between 
the various ``flavours'' of models predicted by N-body simulations 
 since these models only differ in their inner cores
and are therefore inaccessible to weak lensing. Strong lensing analyses
of arcs, sensitive to this inner core might prove to be sufficient 
to constrain the inner parts of the profiles 
but this still remains an open question.   

We also  examined the biases in  mass estimates when an 
incorrect cluster
profile (eg. SIS) is assumed and found  that these biases can be quite
significant ($\sim 70$\%), especially at  low redshifts ($z \sim
0.2$).  At higher  redshifts,  the  bias   is weaker, although   still
substantial at $\sim 50$\%. This redshift dependence was traced to 
the fact that at higher redshifts, images of a given angular extent
correspond to a greater physical extent. 
Weak lensing constrains the mass within the observed area, but in order
to estimate the virial mass (or $M_{1500}$), one must extrapolate to the 
appropriate radius. Using an incorrect profile for this extrapolation will
bias the mass estimate. However, the degree of extrapolation decreases with
increasing redshift, reducing the bias.

Finally, we also showed that the method could extract the correct 
cluster profile parameters although only to within $25$\%. The 
greater errors in this case can be traced to a degeneracy in the 
parameters for a given mass; such a degeneracy would naturally
broaden the likelihood function and increase the errors. The case
for a degeneracy is further strengthed by the fact that the mass estimates
are significantly more robust, being within $10$\%.

This paper represents a first attempt to demonstrate the feasibility
of our method and estimate the reliability of profile determination and the
bias introduced in the mass estimate due to an incorrect profile.
For simplicity, we have left a number of practical aspects 
unaddressed which we now consider. The most relevant issue is the
measurement of the intrinsic ellipticity distribution. We computed this 
distribution self consistently in our simulation. A real world application
must determine this background ellipticity distribution in fields away
from significant weak lensing sources. The Medium Deep Survey (Ratnatunga et.
al 1999) is one such possibility. There are a number of concerns with
this approach such as the evolution of the ellipticity distribution 
with redshift and the effects of the instrument of the ellipticity distribution, 
that still remain unexplored.

A second area that must be incorporated into the method is the redshift 
distribution of the lensed galaxies, since the strength of the lensing 
is dependent on the distance between the galaxy and the lens. There
are two possible approaches to solving this problem within the framework 
of our method. The first is to use photometric redshifts and use those
as a second input parameter in Eq. \ref{eq:like}
\begin{equation}
{\cal L} = - \sum_{i=1}^{N} \log p(\beps_i,z_{i};\delta_i,\theta_i) \,.
\label{eq:like}
\end{equation} 
The second is to convolve a redshift distribution directly into the 
likelihood function and measure its parameters from the data itself.
Eq. \ref{eq:like} would then read
\begin{equation}
{\cal L} = - \sum_{i=1}^{N} \log p(\beps_i;\delta_i,\theta_i,\bf{z_{i}}) \,.
\label{eq:like}
\end{equation}
where $\bf{z_{i}}$ represents the parameters of the redshift distribution.
The generic effect of such an approach would be to broaden the likelihood
function and therefore, increase the errors in the measurements.

Finally, the effect of seeing
on these results must be addressed.
Seeing, as it tends to circularise images, would 
generically reduce the lensing signal and therefore result in an 
underestimate of the mass distribution. The measured 
intrinsic ellipticity distribution would also be affected by seeing,
and this may compensate for the effects of seeing, although a 
definitive answer must wait for further simulations that include seeing. 

In spite of as yet unexplored observational effects relevant to our 
technique, our formalism still calls for an observational test of 
theoretical predictions of cluster profiles. Improvements in
telescopes designs for weak lensing searches ought to make this a
feasible approach in the future.

\section*{Acknowledgements}
N.P. would like to acknowledge J.A.W., to whom
this paper is dedicated, for his 
work and guidance on this project. 
J.A.W. was killed in a accident on June 18, 2000 
after the project was completed but
while this paper was still being written. 
N.P. would also like to thank P. Batra, S. Courteau, V. Petrosian,
G. Squires, M. Strauss, K. Thompson and R. Wagoner for useful 
insights and feedback. N.P. was supported by a Bing Research Scholarship 
from the Physics Department.

{}

\vfill\eject
\appendix

\section{Analytic forms of the Convergence and Shear for Specific Models}

The convergence and shear for a circularly symmetric mass distribution
were given by equations~\ref{eq:kappa} and~\ref{eq:shear_ampl} 
in the main text 
. In this Appendix we derive
specific expressions for the SIS, NFW, and KKPB cluster profiles.
\footnote{The SIS derivation is straightforward; our derivation
for NFW is our own, but has previously been done, albeit
in slightly different form, by Bartelmann 1996. For KKBP,
we present a numerical result as no analytic solution was
found.}

\subsection{{\bf SIS}}

The density law for a softened isothermal sphere with a core radius $r_s$ is
\begin{equation}
\rho(r) = % \frac{\sigma_v^2}{2\pi G}\,\frac{1}{r_s^2+r^2} =
\frac{\sigma_v^2}{2\pi G r_s^2} \left[\frac{1}{1+\left(r/r_s\right)^2}\right]\,.
\label{eq:SIS}
\end{equation}
Making the connection with the formalism developed in \S~2, 
$\eta(x)=(1+x^2)^{-1}$ and $\delta_c=4\sigma_v^2/3(H_0 r_s)^2.$
This form of $\eta(x)$ leads straightforwardly
to the following expressions for $\beta(x)$ and its derivative:
\begin{equation}
\beta(x)=\frac{\pi}{2}\left(\sqrt{x^{2}+1}-1\right)
\end{equation}
\begin{equation}
\beta^\prime(x)=\frac{\pi}{2} \frac{x}{\sqrt{1+x^2}}\,,
\end{equation}
and thus yields the convergence and shear for the SIS model:
\begin{equation}
\kappa(x)=\frac{2\pi\sigma_v^2}{c^2}\,\theta_s^{-1}
\frac{y_{LS}}{y_S}\frac{1}{\sqrt{1+x^2}}\,,
\end{equation}
\begin{equation}
\gamma(x)=\frac{2\pi\sigma_v^2}{c^2}\,\theta_s^{-1}
\frac{y_{LS}}{y_S}
\left(\frac{1}{\sqrt{1+x^{2}}}-\frac{2}{1+\sqrt{1+x^{2}}}\right)\,,
\end{equation}
where $x=\theta /\theta_s$ and $\theta_s=r_s/D_L.$ 

\subsection{{\bf NFW}}

For NFW we have $\eta(x)=x^{-1}(1+x)^{-2},$ and thus
\begin{equation}
\beta(x) = \int_0^x y\,dy \int_0^\infty \frac{dw}{\sqrt{y^2+w^2}\left(1+\sqrt{y^2+w^2}\right)^2}\,.
\label{eq:betaNFW}
\end{equation}
Making the change of variable $\tan\theta=w/y,$ we obtain
\begin{equation}
\beta(x) = \int_0^x y\,dy \int_0^{\pi/2} \frac{\cos\theta\,d\theta}{\left(y+\cos\theta\right)^2}\,.
\end{equation}
We now change the order of integration, and the $y$-integral is readily
done, yielding
\begin{eqnarray*}
\beta(x)=\int_0^{\pi/2} \cos\theta
\ln\left[1+\frac{x}{\cos\theta}\right]\,d\theta \\
 + \int_0^{\pi/2} \frac{\cos^2\theta\,d\theta}{x+\cos\theta}\,-1\,.
\end{eqnarray*}

We now write $\ln(1+x/\cos\theta)=\ln(\cos\theta+x)-\ln(\cos\theta),$
integrate the first term by parts, and then obtain after some manipulation
\begin{eqnarray*}
\beta(x)=\ln x + \int_0^{\pi/2}\frac{d\theta}{\cos\theta+x} \\ 
 + 
\int_0^{\pi/2}
\cos\theta\,\ln(\sec\theta)\,d\theta - 1\,.
\end{eqnarray*}
The second integral above 
has the value $1-\ln 2$ (Gradshteyn \& Ryzhik 1994, Eq. 4.387). Thus,
\begin{equation}
\beta(x)=\ln(x/2) + {\cal F}(x)\,,
\label{eq:bnfw}
\end{equation}
where
\begin{equation}
{\cal F}(x) \equiv \int_0^{\pi/2}\frac{d\theta}{\cos\theta+x}\,,
\end{equation}
and has the following values (Gradshteyn \& Ryzhik 1994, Eq. 2.553):
\begin{enumerate}
\item $x<1$:
\begin{equation}
{\cal F}(x)= \frac{1}{\sqrt{1-x^2}}\,\ln\left(\frac{x}{1-\sqrt{1-x^2}}\right)
\end{equation}
\item $x>1$:
\begin{equation}
{\cal F}(x)= \frac{1}{\sqrt{x^2-1}}\,\cos^{-1} (1/x)
\end{equation}
\item $x=1$:
\begin{equation}
{\cal F}(x)= 1
\end{equation}
\end{enumerate}

Differentiating equation~\ref{eq:bnfw} yields
\begin{equation}
\beta^\prime(x)=\frac{1}{x} + {\cal F}^\prime (x)
\end{equation}
where
\begin{equation}
{\cal F}^\prime (x)=
-\int_{0}^{\pi/2} \frac{d\theta}{(x+\cos\theta)^2}\,.
\end{equation}
Carrying out the integral (Gradshteyn \& Ryzhik 1994, Eq. 2.553) then leads to
\begin{equation}
{\cal F}^\prime (x)= \frac{1}{x^{2}-1} \left [\frac{1}{x}-x{\cal F}(x)\right ]
\end{equation}
and 
\begin{equation}
{\cal F}^\prime (1) = -\frac{2}{3}\,.
\end{equation}
Equations~A10--A18 above enable use to calculate bending angle,
convergence, and
shear for an NFW model of specified $\delta_c$ and $\theta_s,$
by means of equations~\ref{eq:final_lens}, 
\ref{eq:kappa}, and~\ref{eq:shear_ampl} of the main text.

\subsection{{\bf KKBP}}

A KKBP halo has $\eta(x) = x^{-0.2} (1+x^2)^{-1.4},$ and thus
\begin{equation}
\beta(x)=\int_0^x y\,\sigma(y)\, dy\,,
\label{eq:beta1}
\end{equation}
where
\begin{equation}
\sigma(y) =  \int_0^\infty (y^2+w^2)^{-0.1} [1+y^2+w^2]^{-1.4} dw\,.
\label{eq:sigma1}
\end{equation} 
Unlike the SIS and NFW halos, there is no analytical form for
the functions $\beta$ and $\sigma$ for KKBP. 
We have therefore integrated equation~\ref{eq:sigma1} numerically,
and fit the results to a rational function. The following expression
provides a fit accurate to better than 1\%: 
\begin{equation}
\sigma(x) = \frac{1.250+0.965133x}{1+1.18363x+1.06347x^{2}+x^{3}}\,.
\label{eq:sigma2}
\end{equation}
Substitution of this expression into equation~\ref{eq:beta1} 
allows one to obtain an analytic expression for $\beta(x)$ via
integration:
\begin{equation}
\beta(x)=-0.0558209 + \beta_1(x) + \beta_2(x) + \beta_3(x)\,,
\end{equation}
where:
\begin{equation}
\beta_1(x) = 0.126086 \tan^{-1} (0.0605662+0.970072 x)
\end{equation}
\begin{equation}
\beta_2(x) = -0.176946\ln(0.937601+x)
\end{equation}
\begin{equation}
\beta_3(x) = 0.571039\ln(1.06655 + 0.124869x+x^2)\,.
\end{equation}

\section{Determination of Cluster Virial Masses}

The virial mass of a cluster, $M_V,$ is defined to be the mass 
within a sphere whose mean interior density
is larger than the
background density by a particular factor. The radius
of the sphere is called the virial radius, $R_V.$ In an Einstein-de Sitter
universe the factor is $178$; in open and flat low-density
universes the factor is larger. 
For the purposes of this paper, we simply assume the factor is
known, and denote it $\delta_V.$ (In the main body of the
paper we take $\delta_V=200.$) In this Appendix we show how
to calculate $M_V$ and $R_V$ for a given density law.

As usual, we write the halo density profile as
$\rho(r)=\rho_{crit}\delta_c\eta(x),$ where
$x=r/r_s$ and $r_s$ is the characteristic radius.
The mass interior to a radius $R$ is thus
\begin{equation}
M(R) = 4\pi r_s^3 \rho_{crit}\delta_c m(y)\,,
\label{eq:mofr}
\end{equation}
where $y=R/r_s$ and $m(y)$ is the dimensionless
mass profile given by 
\begin{equation}
m(y)=\int_0^y x^2 \eta(x) dx\,.
\label{eq:defmy}
\end{equation}

The condition that the
mean density within $R_V$ 
is $\delta_V$ times the background density is
\begin{equation}
\frac{M(r_V)}{4\pi r_V^3/3} = \delta_V \Omega_M \rho_{crit}\,.
\end{equation}
Combining this with equation~\ref{eq:mofr} yields
\begin{equation}
m(y_V) = \frac{\delta_V}{3 \delta_c} \Omega_M y_V^3 \,,
\label{eq:mdelta}
\end{equation}
where $y_V\equiv R_V/r_s.$
For any given density law $\rho(r)$ and value of $\Omega_M,$
equation~\ref{eq:mdelta} can be solved
for $y_V,$ which in turn yields the virial radius $R_V=y_V r_s,$
and the virial mass through equation~\ref{eq:mofr}.

The integral in equation~\ref{eq:defmy} is readily carried
out for the SIS and NFW density profiles, yielding
\begin{equation}
m_{SIS} (y) = y - \tan^{-1} y 
\label{eq:mysis}
\end{equation}
and
\begin{equation}
m_{NFW} (y) = \ln (1+y) - \frac{y}{1+y}\,.
\label{eq:mynfw}
\end{equation}

It is useful to express the virial mass in terms
of typical halo parameters. From equation~\ref{eq:mofr} one finds:
\begin{equation}
M_V = 3.49\times 10^{14} \left(\frac{r_s}{100\ {\rm kpc}}\right)^3
\left(\frac{\delta_c}{10^5}\right) m(y_V)\,h^{-1} M_\odot\,.
\end{equation}
For example, a KKBP halo with
for $r_s=100$ kpc and 
$\delta_c=10^5$ has $M_V=8.7\times 10^{14}\,h^{-1} M_\odot,$
and $r_V=1.55$ Mpc, for $\delta_V=200.$


\begin{thebibliography}{}
\bibitem[]{} Bahcall, N.A.\ 2000, Phys. Scripta, T85, 32, (astro-ph/9901076)
\bibitem[]{} Bahcall, N.A., \& Cen, R.\ 1993, ApJL, 407, L49
\bibitem[]{} Bahcall, N.A., Fan, X., Cen, R.\ 1997, ApJL, 485, L53
\bibitem[]{} Bartelmann, M., \& Schneider, P.\ 1999, preprint (astro-ph/9912508)
\bibitem[]{} Bartelmann, M.\ 1998, Evolution of Large-Scale Structure: From 
Recombination to Garching, E28 

\bibitem[]{} Bartelmann, M.\ 1996, 313, 697 

\bibitem[Blanchard, Sadat, Bartlett \& Le Dour(2000)]{2000chr..conf..158B} 
Blanchard, A., Sadat, R., Bartlett, J.\ \& Le Dour, M.\ 2000, ASP Conf.\ 
Ser.\ 200: Clustering at High Redshift, 158 (astro-ph/9908037)

\bibitem[Borgani, Rosati, Tozzi \& Norman(1999)]{1999ApJ...517...40B} 
Borgani, S., Rosati, P., Tozzi, P.\ \& Norman, C.\ 1999, ApJ, 517, 40 
(astro-ph/9901017) 

\bibitem[]{} Carlberg, R.\ G.\ et al.\ 1998, Large Scale Structure:  Tracks 
and Traces, 119 

\bibitem[]{} Carlberg, R.\ G., Morris, S.\ L., Yee, H.\ K.\ C.\ \& 
Ellingson, E.\ 1997, ApjL, 479, L19 

\bibitem[Donahue et al.(1998)]{1998ApJ...502..550D} Donahue, M., Voit, G.\ 
M., Gioia, I., Lupino, G., Hughes, J.\ P.\ \& Stocke, J.\ T.\ 1998, ApJ, 
502, 550

\bibitem[Eke, Cole, Frenk \& Patrick Henry(1998)]{1998MNRAS.298.1145E} Eke, 
V.\ R., Cole, S., Frenk, C.\ S.\ \& Patrick Henry, J.\ 1998, MNRAS, 298, 
1145 

\bibitem[Girardi et al.(1998)]{1998ApJ...506...45G} Girardi, M., Borgani, 
S., Giuricin, G., Mardirossian, F.\ \& Mezzetti, M.\ 1998, ApJ, 506, 45 

\bibitem[Gradshteyn \& Ryzhik(1994)]{1994tisp.book.....G} Gradshteyn, I.\ 
S.\ \& Ryzhik, I.\ M.\ 1994, New York: Academic Press

\bibitem[]{} Hoekstra, H., Franx, M., Kuijken, K., \& Squires, G.\ 1998,
ApJ, 504, 636
\bibitem[]{} Kaiser, N., \& Squires, G.\ 1993, ApJ, 404,441

\bibitem[Kaiser, Tonry \& Luppino(2000)]{2000PASP..112..768K} Kaiser, N., 
Tonry, J.\ L.\ \& Luppino, G.\ A.\ 2000, PASP, 112, 76 (astro-ph/9912181)

\bibitem[]{} Kravtsov, A.V., Klypin, A.A., Bullock, J.S., \& Primack,
J.R.\ 1998, ApJ, 502, 48 (KKBP)

\bibitem[Lewis, Ellingson, Morris \& Carlberg(1999)]{1999ApJ...517..587L} 
Lewis, A.\ D., Ellingson, E., Morris, S.\ L.\ \& Carlberg, R.\ G.\ 1999, 
ApJ, 517, 587 (astro-ph/9901062) 

\bibitem[]{} Lewis, A.D., Ellingson, E., Morris, S.L., \& Carlberg, R.G.\
1999, preprint (astro-ph/9901062)

\bibitem[Lewis, Ellingson, Morris \& Carlberg(1999)]{1999ApJ...517..587L} 
Lewis, A.\ D., Ellingson, E., Morris, S.\ L.\ \& Carlberg, R.\ G.\ 1999, 
ApJ, 517, 587

\bibitem[]{} Luppino, G.A., \& Kaiser, N.\ 1997, ApJ, 475, 20
\bibitem[]{} Luppino, G.A., Tonry, J.L., \& Stubbs, C.W.\ 1998, Proc. SPIE,
3355, 469

\bibitem[Miralda-Escude(1991)]{1991ApJ...370....1M} Miralda-Escude, J.\ 
1991, ApJ, 370, 1 (ME91)

\bibitem[]{} Navarro, J.F., Frenk, C.S., \& White, S.D.M.\ 1997, ApJ, 490,
493 (NFW)
\bibitem[]{} Narayan, R., \& Bartelmann, M.\ 1999, in {\em Formation of 
Structure in the Universe,} eds.~A. Dekel \& J.P. Ostriker
(Cambridge University Press) (NB99)

\bibitem[Peebles(1993)]{1993ppc..book.....P} Peebles, P.\ J.\ E.\ 1993, 
Princeton Series in Physics, Princeton, NJ: Princeton University Press

\bibitem[Ratnatunga, Griffiths \& Ostrander(1999)]{1999AJ....118...86R} 
Ratnatunga, K.\ U., Griffiths, R.\ E.\ \& Ostrander, E.\ J.\ 1999, AJ, 
118, 86

\bibitem[Reichart et al.(1999)]{1999ApJ...518..521R} Reichart, D.\ E., 
Nichol, R.\ C., Castander, F.\ J., Burke, D.\ J., Romer, A.\ K., Holden, 
B.\ P., Collins, C.\ A.\ \& Ulmer, 
M.\ P.\ 1999, ApJ, 518, 521 (astro-ph/9802355)

\bibitem[Sadat, Blanchard \& Oukbir(1998)]{1998A&A...329...21S} Sadat, R., 
Blanchard, A.\ \& Oukbir, J.\ 1998, A\&A, 329, 21

\bibitem[]{} Schneider, P., \& Seitz, C.\ 1995, A\&A, 294, 411
\bibitem[]{} Seitz, C., \& Schneider, P.\ 1995, A\&A, 297, 287
\bibitem[]{} Seitz, C., \& Schneider, P.\ 1996, A\&A, 305, 383
\bibitem[]{} Seitz, C., \& Schneider, P.\ 1997, A\&A, 318, 687
\bibitem[]{} Seitz, C., Schneider, P., \& Bartelmann, M.\ 1998, A\&A, 337, 325

\bibitem[Smail et al.(1997)]{1997ApJ...479...70S} Smail, I., Ellis, R.\ S., 
Dressler, A., Couch, W.\ J., Oemler, A.\ J., Sharples, R.\ M.\ \& Butcher, 
H.\ 1997, ApJ, 479, 70 

\bibitem[Squires \& Kaiser(1996)]{1996ApJ...473...65S} Squires, G.\ \& 
Kaiser, N.\ 1996, ApJ, 473, 65

\bibitem[Squires et al.(1997)]{1997ApJ...482..648S} Squires, G., Neumann, 
D.\ M., Kaiser, N., Arnaud, M., Babul, A., Boehringer, H., Fahlman, G.\ \& 
Woods, D.\ 1997, ApJ, 482, 648

\bibitem[]{} Tran, K.-V. H., Kelson, D.D., van Dokkum, P., 
Franx, M., Illingworth, G.D.,
\& Magee, D.\ 1999, ApJ, 522, 39 

\bibitem[Tyson \& Seitzer(1988)]{1988ApJ...335..552T} Tyson, J.\ A.\ \& 
Seitzer, P.\ 1988, ApJ, 335, 552 

\bibitem[Tyson(1988)]{1988AJ.....96....1T} Tyson, J.\ A.\ 1988, AJ, 96, 1 

\bibitem[]{} Tyson, J.A., Kochanski, G.P., \& Dell'Antonio, I.P.\ 1998,
ApJ, 498, L107

\bibitem[Tyson, Wenk \& Valdes(1990)]{1990ApJ...349L...1T} Tyson, J.\ A., 
Wenk, R.\ A.\ \& Valdes, F.\ 1990, ApJL, 349, L1

\bibitem[]{} Valdes,\ F., 1982, Faint Object Classification and Analysis System,
NOAO document

\bibitem[van den Bosch \& Swaters(2000)]{2000astro.ph..6048V} van den 
Bosch, F.\ C.\ \& Swaters, R.\ A.\ 2000,\ Submitted 
for publication in AJ (astro-ph/0006048)

\bibitem[White, Efstathiou \& Frenk(1993)]{1993MNRAS.262.1023W} White, S.\ 
D.\ M., Efstathiou, G.\ \& Frenk, C.\ S.\ 1993, MNRAS, 262, 1023 

\bibitem[Willick(2000)]{2000ApJ...530...80W} Willick, J.\ A.\ 2000, ApJ, 
530, 80 

\bibitem[Wilson, Cole \& Frenk(1996)]{1996MNRAS.280..199W} Wilson, G., 
Cole, S.\ \& Frenk, C.\ S.\ 1996, MNRAS, 280, 199

\bibitem[]{} Wu, X.P., Chiueh, T., Fang, L.Z., \& Xue, Y.J.\ 1998,
MNRAS, 301, 861
\end{thebibliography}
\end{document}